\pdfoutput=1

\documentclass[oneside,11pt]{Latex/Classes/PhDthesisPSnPDF} 

\newcommand{\figuremacroW}[4]{
	\begin{figure}[htbp]
		\centering
		\framebox{
		\includegraphics[width=#4\textwidth]{#1}
		}
		\caption[#2]{\textbf{#2} - #3}
		\label{#1}
	\end{figure}
}

\usepackage[T1]{fontenc}
\usepackage[utf8]{inputenc}
\usepackage{graphicx}
\usepackage{listings}
\usepackage{algorithmic}
\usepackage{algorithm}

\ifpdf  
    \pdfcatalog { /PageMode (/UseOutlines)
                  /OpenAction (fitbh)  }
\fi

\title{Visual definition of procedures for automatic virtual scene generation}

\ifpdf
  \author{\href{mailto:drazen.lucanin@gmail.com}{Dražen Lučanin}}
  \collegeordept{\href{http://www.fer.hr}{Faculty of Electrical Engineering and Computing}}
  \university{\href{http://www.unizg.hr}{University of Zagreb}}

\else
  \author{Dražen Lučanin}
  \collegeordept{Faculty of Electrical Engineering and Computing}
  \university{University of Zagreb}
\fi

\degree{Master of Science in Computing}
\degreedate{Zagreb, June 2011}

\begin{document}


\renewcommand\baselinestretch{1.2}
\baselineskip=18pt plus1pt


\maketitle  


%
%
%






\frontmatter

\begin{dedication} 

To my parents, Damir and Jasminka, for their wholehearted support throughout my life and studies.

\end{dedication}


\begin{acknowledgements}      

The author whishes to express his gratitude to Professor Helmut Hlavacs, ao.univ.prof.dr. at the Faculty of Computer Science, University of Vienna and Professor Domagoj Jakobović, doc.dr.sc. at the Faculty of Electrical Engineering and Computing, University of Zagreb, for mentoring this research and their help in writing the thesis.

This thesis was written and the accompanying work done at the Faculty of Computer Science, University of Vienna, as part of the Erasmus student exchange programme.  

\end{acknowledgements}



\setcounter{secnumdepth}{3} 
\setcounter{tocdepth}{3}    
\tableofcontents            


\listoffigures	



%
%
%


\mainmatter





\chapter{Introduction}

\ifpdf
    \graphicspath{{1_introduction/figures/PNG/}{1_introduction/figures/PDF/}{1_introduction/figures/}}
\else
    \graphicspath{{1_introduction/figures/EPS/}{1_introduction/figures/}}
\fi


Creating content in the modern era of multimedia and interactive virtual applications is a difficult and lengthy task, usually tackled by a whole group of experts. Procedural content generation or modeling is an approach which aims to make this more efficient by requiring only a description of the content and generating the content from it. Historically, methods such as L-systems \cite{prusinkiewicz_algorithmic_1991} have been used for this process, but they require a good knowledge of programming, often relying on editing rules in textual files using a complicated syntax and are therefore only adopted to people with specialized knowledge in that field.

Taking into consideration the quality of modern graphical user interfaces (GUIs), it somehow seems that this approach could be improved and brought to a wider profile of users. Can the classical textual programming language environments be replaced with more modern graphical applications? There are few key points in discussing positive aspects of developing visual programming languages (VPLs) in general.

An additional emphasis in this thesis will be put on the usage of a VPL for the task of defining procedures that can be used for automatic virtual scene generation. Virtual scenes are a big part of virtual content and can be very complex and therefore hard for manual creation, which makes their procedural generation an especially interesting problem.

\section{The strengths of visual programming}
\label{sec:strengths}

According to Knuth's definition in \cite{knuth_art_1997}, an algorithm has five important features:
\begin{itemize}
	\item finiteness -- an algorithm has to terminate within a finite amount of time
	\item definiteness -- every step of the algorithm must be precisely and disambigously defined
	\item input -- every algorithm needs to have some data given to it as input
	\item output -- when an algorithm terminates it should provide some data as output
	\item effectiveness -- an algorithm must effectively perform its task in a minimal amount of time.
\end{itemize}

The most important feature we will examine in this context is definiteness -- the steps of an algorithm have to be well defined, disambiguous. We can infer from this that to implement an algorithm, a computer program also requires definiteness.

In classical programming languages (such as C, defined in \cite{kernighan_c_1988}) the programmer writes his program down in a textual file, compiles it to machine code and is able to execute it. Textual files can consist of any imaginable text -- from Shakespeare's dramas to random gibberish. Well defined classical programs are a very small subset of text. In classical programming languages, definiteness is checked mostly through programming language \textit{syntax} (it is also achieved in part through semantic and runtime checks at later stages of program compiling and running, but that will not be discussed here). Syntax is essentially a set of rules that constraint text that qualifies as valid -- more details can be found in \cite{hopcroft_introduction_2007}. The programmer tries to describe an algorithm he has imagined by writing commands in some syntax and getting notifications about their correctness afterwards. Fig. \ref{VPL-CPL-diagram-combo} illustrates this approach, in which the user manually transforms his logic into syntax. This is in its essence a very black-box approach, i.e. the programmer receives only a posteriori notifications about the program's definiteness. Of course, there are some features that help in the process, like dynamic type checking or auto-completion which save time, but they do not let the programmer fully forget about the syntax.

\figuremacroW{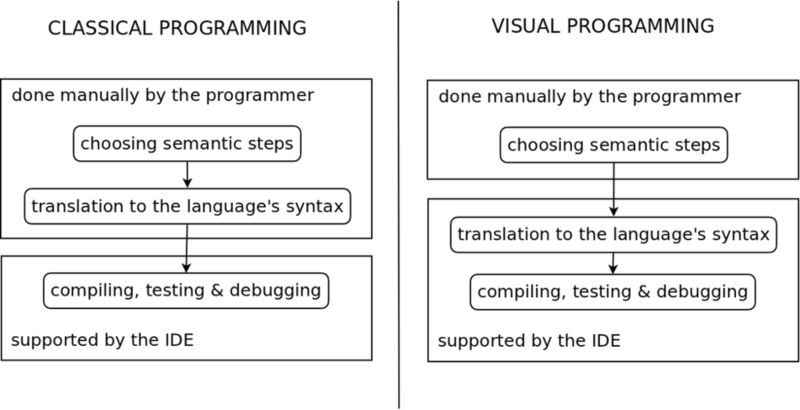}{Classical vs. visual programming work distribution}{A diagram showing the distribution of work between the programmer and the IDE in classical and visual programming}{0.9}

In VPLs the programmer does not input text, but uses modern GUI capabilities to drag \& drop blocks of commands, connections between them etc. to describe an algorithm. This is by itself a reduction of the possible constructs that can be made by a programmer -- only a couple of symbols (defined by the language designer) are available, in contrast to the whole alphabet (and more) in classical programming. \textbf{The choices a programmer has to make when defining programs are limited to the ones that satisfy the language syntax only.} An illustrative way to describe this would be to say that in VPLs the keyboard has been altered to contain only the keys vital to programming (sort of like the old Spectrum computers \cite{_zx_????}). Another important aspect is \textbf{the chance for preeminent notifications} -- notifying the programmer about the possible constraint breeches (syntactic or semantic errors) while he is still dragging an element in the GUI -- possibly saving the time of making a mistake. This way a programmer is guided to satisfy constraints as proposed in \cite{burnett_scaling_1995}, unlike syntax errors which just notify the user when he is already breached one. Finally there is \textbf{the visual overview aspect} -- unlike classical programs where we have to first focus on specific letters and read them to discover what type of algorithm pattern we're dealing with, in visual programming we have the ability to recognize certain patterns in a purely visual way (by recognizing circles, squares, lines, colour etc.), leading to an examination of the meaning of some program in a more top-down manner.

All in all, an attempt is made in VPLs to \textbf{shift more of the work from the programmer to the integrated development environment (IDE)}. As illustrated in Fig. \ref{VPL-CPL-diagram-combo}, the task of syntax generation is done automatically -- allowing the programmer to focus more on the abstract notions of algorithm design.

To summarize, the benefits of this shift are:
\begin{itemize}
  \item \textbf{easy syntax rule obeying} -- only the freedom of creating programs is given to the programmer
  \item \textbf{a better notification system} -- due to the fact that the IDE has more knowledge about the program and the programmer's actions at an earlier moment than classical programming IDE
  \item \textbf{visual representation can be easier to grasp}
\end{itemize}

All of these elements could in theory be achieved in an IDE and presented to the user as a rich, context-avare VPL. Such an environment would make the task of programming a user-friendlier experience and help a broader audience of previously end-users to start performing it.

\section{Introduction to procedural scene modeling}
\label{sec:introduction.procedural}
Procedural scene generation or modeling is a way of creating scenes by simply defining a sequence of instructions that manipulate units of small complexity offered by the graphical engine (such as polygons, textures, simple premade objects). When the computer follows this set of instructions the result is a whole virtual scene being created -- more complex than the individual units used to compose it.

Procedural modeling is in a way similar to visual programming as it presents the user only with a smaller set of possible decisions and the hard work of processing these decisions and making them compatible with the lower layer (the graphical engine in this case) is performed automatically.

The good aspects of procedural generation are:
\begin{itemize}
  \item it saves space -- a procedure describing the scene will take much less physical memory than stating the geometrical representations of all the objects which populate it
  \item it saves time -- which follows quite directly from the first statement, as an artist does not have to manually specify each of the objects that are inside the scene, but instead does that on the meta-level of a procedure.
\end{itemize}

Possible applications for procedural scene generation can be found in basically any kind of virtual world (computer games, simulations etc.) -- to succeed in generating content faster. Another potentially interesting target is dynamical scene creation -- a scene could be generated differently each time the user visits it. In the field of computer games, this could add to the game's replay value or even be used to create experimental games with altered laws of physics.


\chapter{Related work}

Visual programming has been researched extensively in the past. In the following section a short overvies of the state of the art is presented.

After that the recent advances in procedural scene modeling will be examined.

\section{Advances in visual programming}
\label{sec:related_work.visual}

In \cite{burnett_scaling_1995} a general overview of the visual programming strengths and weaknesses is examined with a lot of examples of how certain languages cope with these opportunities and challenges. The two interesting concepts are mentioned:
\begin{itemize}
  \item \textbf{static representation} -- is a graphical notation used to present a program at rest sufficient to understand its meaning
  \item \textbf{effective use of computer display} -- showing only the information important to the user at that given time, considering the context
\end{itemize}

These two factors are something important if a VPL is to scale well to a project of a larger scope. By keeping them in mind as guidelines while designing a VPL, one could get closer to the usability classical programming currently offers.

\subsection{Dataflow VPLs}

In functional programming, functions are in focus, i.e. the main flow of the program represents data going through a sequence of functions. The order in which these functions will be executed is only partially controlled by function chaining and other than that the programmer does not concern himself with this matter -- the interpreter, a shell that runs the functional program will decide on the order of the independent functions. More about functional programming on the example of LISP can be read in \cite{_programming_1964}.

Hand in hand with the logic of functional programming go dataflow VPLs, which are basically their visualisation. Blocks represent functions and the connections represent ``pipes'' through which the data flows. Dataflow VPLs are a big group in visual programming. They are described from an abstract level and put into context with functional programming in \cite{johnston_advances_2004}.

Because the input and output interfaces of each block are determined as the domain and codomain sets of a coresponding function, it is possible to precisely determine what sort of data is contained in a certain pipe. This is useful for contextually relevant nottifications (for example suggesting appropriate blocks that the pipe can be connected to), which is a strong advantage of visual programming, as mentioned in \cite{burnett_scaling_1995}.

\subsubsection{Domain-specific dataflow VPLs}

Perhaps it is due to these positive features, that dataflow VPLs are especially attractive as domain-specific programming environments, adapted to end-users specialized in some work, without a background in programming. In any case, there exists a great number of domain-specific dataflow VPLs in all kinds of fields of expertise.

LabView \cite{_labview_????} and Simulink \cite{_simulink_????} can be mentioned as very popular representatives of dataflow VPLs. They are mostly aimed for system modeling and development with a strong emphasis on a modular, engineering approach through encapsulation.

Some other examples of domain-specific dataflow VPLs and their domains can be seen in Table \ref{tab:dataflow}.

\begin{table}
\centering
\begin{tabular}{ | p{0.3\textwidth}| p{0.6\textwidth} | }
\hline
Domain & VPL\\
\hline\hline
engineering & LabView \cite{_labview_????}, Simulink \cite{_simulink_????} \\\hline
data mining, web crawling & Rapidminer, Yahoo Pipes \cite{loton_working_2008} \\\hline
architecture & Grasshoper \cite{_grasshopper_????} \\\hline
spatial analysis & ProVal 2000 \cite{_proval_????} \\\hline
image analysis & Harpia \cite{_harpia_????} \\\hline
audio synthesis & SynthMaker \cite{_synthmaker_????} \\\hline
visualisation, animation & Quartz Composer \cite{_working_????}, Toon Boom Animate Pro \cite{_animate_????}, VSXu \cite{_vsxu_????} \\\hline
robotics & Lego Mindstorms NXT \cite{_lego_????} \\\hline
game creation & SourceBinder \cite{_sourcebinder_????} \\\hline
\end{tabular}
\caption[Domain-specific dataflow VPLs]{\textbf{Domain-specific dataflow VPLs} - An incomplete list showing some of the dataflow VPLs that exist for certain professional domains.}
\label{tab:dataflow}
\end{table}

\subsection{Flowcharts}

As a contrast to dataflow VPLs based on functional programming, there is a family of VPLs visually closer to flowcharts and imperative programming. An example imperative programming language is C \cite{kernighan_c_1988}. In imperative programming the programmer defines each step of a program as a command and the order of these commands strictly determines the order in which the computer executes them. This command flow can be controlled using mechanisms such as loops or branches. In a flowchart, a visualisation of an imperative program, we may say that it is the ordering, arranging the sequence of commands (blocks) and not the data itself that is the center of focus. The data somehow remains in the background, stored in variables and fetched when needed.

Even though there appears to be more dataflow than flowchart VPLs (probably due to it being more easy to grasp for newcomers -- no need to study abstract notions such as variables and flow control), professional programmers tend to use imperative languages more as can be seen in Fig. \ref{programming_language_popularity} (results taken from \cite{_compare_????}) -- Python \cite{rossum_introduction_2011}and C \cite{kernighan_c_1988} as mostly imperative languages have a much bigger user base than LISP \cite{_programming_1964}, which is a functional language. It seems from these statistics that if a VPL is to scale well, be adaptable to bigger projects as defined by Burnett et al. \cite{burnett_scaling_1995}, it should be imperative in its basis. Because of this, flowchart VPLs were chosen as the focus in this thesis.

KTechlab \cite{_ktechlab_????} is a flowchart VPL used to describe hardware components. Scratch \cite{_scratch_????} is a good example of a WHILE language flowchart VPL that has taken the role of education through fun and interactive app programming. One drawback in Scratch is that it does not present the user with the source code generated from his flowchart, thus not encouraging gradual transition to classical programming as the user gets more experienced. Alice \cite{_alice:_1995} is another example of an educational VPL that lets the user build interactive graphical applications, while teaching him basic programming concepts.

\figuremacroW{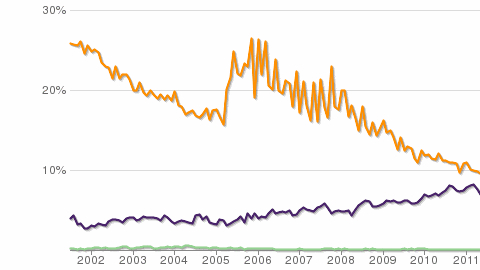}{Programming language popularity}{The count of monthly commits made in a certain programming language in open source projects during the last decade. Orange is C, purple is Python and green is LISP.}{0.7}

\section{Advances in procedural scene modeling}
\label{sec:related_work.procedural}

Since the idea of being able to generate objects or even whole scenes from simple procedures has always been enticing - both from the memory-saving and the artist-time-saving perspective, there has been a lot of research in this field.

\subsection{Grammar-based general approach}

The most-used approach for procedural modeling is using formal grammars
\cite{hopcroft_introduction_2007}, drawing from the same methods used for programming language syntax parsing (the direction is a bit different in procedural modeling, though -- sequences of tokens are randomly generated to satisfy the production rules, unlike programming languages, where existing sequences are parsed to determine the order of applied productions).

Two variants exist -- L-systems \cite{prusinkiewicz_algorithmic_1991}, which operate on strings, and shape grammars \cite{stiny_introduction_1980}, which operate directly on shapes. The basic idea is the same, though -- the user expresses the desired procedure by defining a set of grammar rules -- productions from one sequence of tokens to a different sequence of tokens. The user is, therefore, required to know the theory of formal grammars, production rules, syntax etc.

\subsubsection{L-systems}

Using L-systems -- an algorithmic, procedural generation method for virtual object generation, real-world objects can be replicated. This method relies on randomly choosing productions in formal string grammars \cite{hopcroft_introduction_2007} while keeping predefined global constraints satisfied. The resulting strings represent graphical models -- they encode them. From these encodings the actual graphical models can be drawn in a scene. The way the user affects end-models is by defining the grammar and the constraints. This method was proposed in \cite{prusinkiewicz_algorithmic_1991} for plant modeling.

L-systems were later adapted to building and city generation, as proposed in \cite{mueller_procedural_2006} -- they achieved good results, which can be seen in CityEngine, a commercial system for procedural city generation that implements the methods proposed in \cite{mueller_procedural_2006,parish_procedural_2001}
 
\subsubsection{Shape grammars}

A similar approach was developed by Stiny as shape grammars in \cite{stiny_introduction_1980}. This is also a formal grammar-based approach, but it operates directly on shapes, unlike L-systems which operate on strings.

Shape grammars have also, with slight variations, been used for procedural building modeling in \cite{wonka_instant_2003,mueller_procedural_2006}

\subsection{Often-used algorithms}

\subsubsection{Geographical data consideration}

A method to make an urban scene more realistic is taking sample geographical data into consideration in the modeling process. This approach was proposed in \cite{parish_procedural_2001} and is used in CityEngine in a way that the user can provide various colour-coded map images as input, such as:
\begin{itemize}
  \item a topographic map
  \item a population density map
  \item \ldots
\end{itemize}

The maps can then be analysed and the data taken into consideration during the generation process (for example, higher buildings could be built in areas with higher population densities).

\subsubsection{Road generation}

Both in CityEngine \cite{parish_procedural_2001} and Citygen  \cite{kelly_interactive_2007}, a parallel growth algorithm has been used for road generation, which allows for natural-looking street networks, while being computationally efficient.

In \cite{kelly_interactive_2007}, Kelly also proposes a sampling algorithm with randomness and constraints for mapping roads to terrain in a more natural way, not simply following a straight path. He also proposed several strategies for coping with varying terrain altitudes:
\begin{itemize}
  \item minimum elevation
  \item minimum elevation difference
  \item even elevation difference
\end{itemize}

The benefits of these methods are the roads which seem to integrate well with the terrain. This results in more convincing scenes for the user.

\subsubsection{District generation}

After generating a road network, the enclosed regions can be used as city cells -- areas for buildings and parks, or districts on a wider scale (areas enclosed by bigger roads). A way to detect these regions can be by using the Minimum Cycle Basis algorithm as proposed in \cite{kelly_interactive_2007}.

Another approach that might be used for district generation is by using Voroni diagrams, a method of generating natural-looking cells, as explained in \cite{ebert_texturing_2002}.

\subsection{Post-processing of the scene}

Some of the tools available for procedural generation of scenes offer a possibility of post-processing the scene -- editing it after it has been created. This allows the user to change some details, play with the placement of objects etc. so that the scene can better fit his needs. This is implemented through a what-you-see-is-what-you-get (WYSIWYG) editor of the scene, where the user can move elements around and see what the scene looks like at the same time through real-time rendering.

The CityEngine \cite{parish_procedural_2001} and Citygen \cite{kelly_interactive_2007} systems offer WYSIWYG editors for urban scenes. In a prototype presented in \cite{lipp_interactive_2011}, transformation and merging operators for topology preserving and topology changing transformations based on graph cuts are proposed. Such techniques allow for more intuitive calibration of the final scene through drag-and-drop interaction and are a good combination of procedural and manual modeling of the scene.


\chapter{Goals}
\label{sec:goals}

The goal of this thesis is to present a visual programming language as a tool for the definition of procedures for automatic scene generation. The methods for building a visual programming language are a continuation of the work done in \cite{luanin_visual_2011} and procedural generation is researched as a possible domain for the visual programming language. 

\section{Visual programming goals}
\label{sec:goals.visual}

In this thesis, the methods required to create a VPL for editing and executing flowcharts with an additional ability to generate classical source code from them will be presented. The goal programming language (in which the code is to be generated) is a structured WHILE language (such as Python \cite{rossum_introduction_2011}), to offer good readability. This would give the user another educational step between Scratch \cite{_scratch_????} and classical programming, where he would be able to draw a flowchart and generate good quality, readable code from it and execute it. The code readability would let him study and understand the syntax created from his flowcharts.

These methods will be explained on an example of vIDE, a VPL that was built to fulfil the requirements. The motive behind vIDE is to lower the barrier of learning programming for children as well as for other experts who do not know a programming language syntax, but need to implement certain algorithms.

Firstly, \textbf{a flowchart editor} is needed and the problem of synchronizing a graphical representation with a model will be addressed. Secondly, we will cover the problem of \textbf{generating code in a certain classical programming language from this model}.

An interesting aspect of code generation that will be explored is the internal representation of an algorithm in a GOTO manner, the most similar in logic to a flowchart, and its transformation to a WHILE language data structure. GOTO and WHILE languages are formally defined in \cite{saabas_compositional_2006}. In short -- GOTO languages use a GOTO instruction to jump anywhere in the program, while WHILE languages have WHILE and IF instructions to conditionally execute a certain block (zero, one or multiple times) thus adding more structure and making the flow more predictable. It will be shown that the transformation from a GOTO language to a WHILE language representation is a very practical way of building a flowchart VPL because of the difference between the VPL conceptual model and the output classical programming language.

\section{Procedural modeling goals}
\label{sec:goals.procedural}

Unlike CityEngine \cite{parish_procedural_2001} and similar solutions based on L-systems which rely on editing formal grammars \cite{hopcroft_introduction_2007}, the approach we wish to pursue is to allow the user to \textbf{define the scene modeling procedure in a program} -- without going to the meta level of programming language grammars. Diversity can be achieved by simply using random numbers in the program, instead of randomly generating ``static'' programs (which define scenes deterministically, without any randomness) as is the case with L-systems. This will result in the user having to know only simple programming. He would not need to know the theory of grammars, productions, tokens, final symbols etc. which is required to describe a modeling procedure in CityEngine and other similar systems.

In this thesis, the methods of encapsulating procedural modeling algorithms in a library and exposing them to the user through an application programming interface (API) will be presented. As a proof-of-concept example, an urban scene will be targeted, but similar principles might apply to different scenes as well. The underlying graphics engine used to actually render the scene objects is Ogre \cite{_ogre_????}, an open source graphics rendering engine.

Furthermore, to reach an even broader audience, the programming language that is to be used for defining the scene modeling procedure is to be a \textbf{visual programming language}. The benefit of such a choice is two-way:
\begin{enumerate}
  \item procedural modeling is simpler for end-users
  \item vIDE gets a domain, possibly attracting users  -- similar to the way Scratch \cite{_scratch_????} can be used to create games and animations to encourage children to learn programming concepts through play
\end{enumerate}

It is also important to take into account that the procedural modeling library is to be used from a visual programming language and design it accordingly. For example, since diagrams take up a lot of space and the screen reel becomes precious in visual programming \cite{burnett_scaling_1995}, the commands needed to define the scene modeling procedure should work in a simple form, without any redundancy, but be configurable to specify details if necessary -- that would also make concentrating on the meaning of the procedure easier.

\section{Organisation of the thesis} 

The thesis will deal separately with the two matters:
\begin{enumerate}
  \item the methods of building a VPL, with vIDE as an example
  \item the methods of procedural scene generation and how they can be adapted to visual programming
\end{enumerate}

In Chapter \ref{sec:vIDE} the methods used to create a VPL are presented. In Section \ref{sec:system_architecture} a system architecture of vIDE will be explained so that certain modules can be referenced later in a more clear way. The problem of diagram-to-model mapping is discussed in Section \ref{sec:flowchart_editor}. The motives and mechanics behind the GOTO-to-WHILE model-to-model transformation used in vIDE is given in Section \ref{sec:goto-while}. Finally, the last step of generating code is shortly explained in Section \ref{sec:model-code}.

In Chapter \ref{sec:procedural} the methods of procedural scene generation are covered with a special emphasis on visually defining these procedures. In Section \ref{sec:system_architecture.libprocedural} the library libprocedural that was developed to implement the proposed methods is presented with its system architecture -- the library's interface (Section \ref{sec:libprocedural.python_interface}) and its communication and deployment alongside the Ogre graphics engine at the back (Section \ref{sec:librpocedural.ogre}) and vIDE at the front end (Section \ref{sec:libprocedural.deployment}). Section \ref{sec:libprocedural.api} analyses the library's interface -- the philosophy used to design it in accordance with visual programming in mind (Section \ref{sec:libprocedural.api.simple_by_default}), and the overview of how one would use it programmatically (Sections \ref{sec:libprocedural.api.exposed} and \ref{sec:libprocedural.api.analysis}). The methods used inside the library are presented in Section \ref{sec:libprocedural.modules}.

Chapter \ref{sec:results} presents the results after implementing the methods proposed in the thesis in vIDE and libprocedural. In Section \ref{sec:results.vide} we examine the capabilities of vIDE -- a simple example of usage, the special cases of constraint satisfaction and a useful feature of knowledge inferring from the model. Section \ref{sec:results.libprocedural.programmatically} introduces the possibilities of what could be created using libprocedural by simply invoking it from a classical programming language and gives a side-by-side comparison of the high-level and low-level interfaces that can be used. Section \ref{sec:results.procedural} shows the capabilities of libprocedural in combination with vIDE on a series of use-case examples, gradually covering most of the implemented procedural generation methods.

Chapter \ref{sec:conclusion} places the results achieved through vIDE and libprocedural in the context of the initial objectives and related work in visual programming and procedural scene generation and summarizes the advantages and possible improvements of the used techniques.

\chapter{A visual programming language for drawing and executing flowcharts}
\label{sec:vIDE}

\ifpdf
    \graphicspath{{2/figures/PNG/}{2/figures/PDF/}{2/figures/}}
\else
    \graphicspath{{2/figures/EPS/}{2/figures/}}
\fi


In order to be fully usable as a programming language, a VPL must allow the programmer to:
\begin{enumerate}
  \item express an algorithm's logic in some manner
  \item be able to translate this logic into executable machine instructions
\end{enumerate}

Next, the methods used in implementing our VPL, vIDE, to achieve this will be analyzed.

\section{System architecture -- vIDE}
\label{sec:system_architecture}

The system architecture of vIDE, the prototype VPL presented in this thesis, is illustrated in Fig. \ref{vIDE-arhitektura}.

\figuremacroW{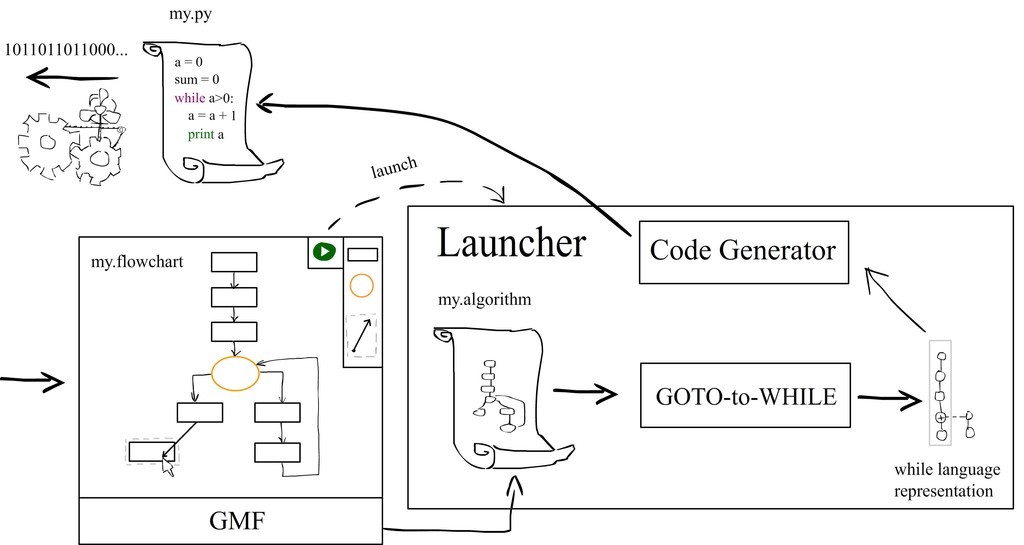}{vIDE architecture}{A graphical overview of the vIDE system architecture}{0.9}

In a very general sense vIDE achieves the basic programming language functions through:
\begin{itemize}
  \item a\textit{ flowchart editor} -- where the programmer visually defines his logic
  \item a \textit{launcher} (a compiler of sorts) -- which transforms the user's flowchart into source code that can be executed
\end{itemize}

The first step to achieve this is to allow the user to draw a flowchart. This was implemented using the Eclipse Graphical Modeling Framework (GMF), a framework for creating diagram editors. As seen in Fig. \ref{vIDE-arhitektura}, an algorithm model is kept in sync with the flowchart that the user draws. GMF does this automatically according to a predefined mapping model). Further information about GMF can be found in \cite{gronback_eclipse_2009}.

The flowchart can be \textit{launched} through the GUI when the user is happy with it. This action initiates a transformation of the algorithm model represented in a GOTO language manner to an equivalent WHILE language style representation (this is required because a WHILE representation can not be created by GMF directly from the vIDE flowchart graphical notation -- this will be further discussed in Section \ref{sec:goto-while}).

From the WHILE language data structure, a concrete textual syntax can easily generate code in any programming language that relies on WHILE and IF commands to control the program flow (for example C \cite{kernighan_c_1988} or Python \cite{_python_????,rossum_introduction_2011}). In vIDE, a Python syntaxer module is called to generate a Python script as the output program, equivalent to the user's flowchart.

Once the output program is generated, the user can study the source code to learn its syntax or simply run it and observe the effects of the flowchart he drew.

In the next few sections the vital methods needed to create a VPL able to generate code will be examined using vIDE as an example implementation.

\section{The flowchart editor -- mapping a diagram to a model}
 \label{sec:flowchart_editor}
 
In every VPL some sort of diagram editor is needed that can be map a graphical notation to a model. A flowchart editor in vIDE was created using GMF, which uses a nice abstract way of describing the diagram editor. In this part it does not really matter what sort of a diagram we are building (flowcharts are just special types of diagrams). The way GMF works is that it requires several models which define the diagram editor's behaviour. Using these models, the diagram editor Eclipse plug-in can then be generated. Methods for building the flowchart editor will be given here in relation to the GMF architecture, but these or similar modules would be required to create a diagram editor utilizing any other technology as well.

The models needed to create a diagram editor, that is, their implementations for vIDE through GMF are:

\begin{itemize}
  \item \textbf{A graphical definition} -- describing the graphical elements that the user will see in his diagram. In vIDE that would be the block, branch and connection elements.
  \item \textbf{A tooling definition} -- definitions of available tools for drawing the diagram. One tool is defined in vIDE for every graphical element.
  \item \textbf{An Eclipse Modelling Framework (EMF) model} -- this is basically the goal data structure synchronized with the diagram as the user edits it. In vIDE a simple GOTO-like algorithm data structure is used and its class diagram can be seen in Fig. \ref{vIDE-class_diagram}.
  \item \textbf{A mapping model -- }this is the heart of the diagram editor description, it defines how to map elements from the graphical definition to an EMF model. GMF uses this definition to synchronize the diagram and the model automatically. If an algorithm model uses a GOTO representation (as is the case in vIDE), this mapping is pretty straightforward -- a block element maps to a block class, branch to branch etc.
\end{itemize}

Such modular design allows for good control of the resulting diagram editor. By implementing all of the mentioned models, vIDE can provide a working flowchart editor through the Eclipse IDE (with additional benefits that will be examined in Chapter \ref{sec:results}).

\figuremacroW{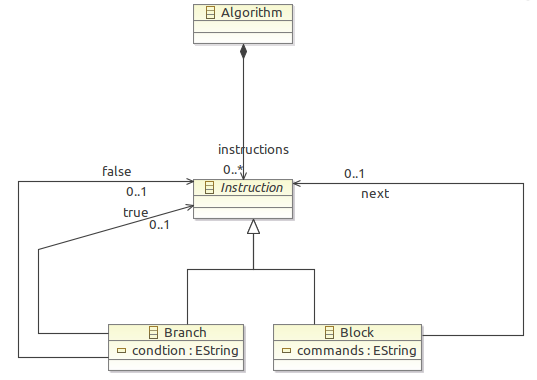}{vIDE class diagram}{Algorithm EMF model used in vIDE - a GOTO-like data structure generated from the flowchart}{0.8}

\section{ An algorithm for the GOTO-to-WHILE transformation}
\label{sec:goto-while}
 
After an algorithm model has been created, a program could be generated disambiguously. The problem is that the algorithm model is described in a GOTO language manner and we want to generate a program in a WHILE language, motivated by the goals stated in Section \ref{sec:goals.visual}. To solve this, a model-to-model transformation is necessary. First, we will examine the reasons for organizing the data structures in such way more deeply and then the transformation algorithm itself will be given.

\subsection{Reasons for using a GOTO language}

For a representation of algorithms in flowcharts, a GOTO language seems more natural, because of a single connection (or we might say flow) going out of every block.

Other possible graphical representations were explored - with special graphical elements for WHILE and IF commands, but these approaches were dropped, because there would have to be multiple outgoing flows and additional explanations of these elements' behaviours would be necessary, which would complicate the usage of vIDE.

\subsection{Reasons for NOT using a GOTO language}
 
It was explained in the last subsection that GOTO is a suitable language for the graphical representation of an algorithm. For the generated programs, on the other hand, WHILE-languages should generally be used instead.
 
In his open letter, Dijkstra expressed his opinion against using GOTO instructions in programming \cite{dijkstra_go_1979}, while Knuth said in \cite{knuth_structured_1974} that they can be useful at times. Generally, we can conclude that GOTO commands can be used carefully (an example of proper usage of GOTO commands in modern languages are exceptions), but WHILE, IF and similar commands should be used to create structured programs. Structured programs are defined in \cite{knuth_structured_1974}.

One of the good sides of structured programming is readability -- if a programmer wants to see what was generated from his flowchart, it would be much more readable (not to mention educational) if loops would be interpreted as WHILE commands and normal branches as IF commands (instead of everything mapping to simple GOTO commands).

A WHILE language representation of the flowchart might be useful to have in a data structure for visual purposes as well, because the usage of structured programming would allow features such as graph folding (letting the programmer hide unnecessary nodes manually or let the IDE do it for him using context-aware technologies such as \cite{_mylyn_????}).

Python was chosen for an output programming language in vIDE because of its simplicity and readability, so another practical reason for not generating code using GOTO commands in vIDE in specific is that Python does not have a GOTO command in its language.

\subsection{Combining the two - the constrained GOTO}
\label{sec:constrained_goto}

When a pure GOTO language representation of an algorithm is stored in a data structure it is clearly a \textbf{graph} -- it can contain loops to already executed commands. When an algorithm represented in a structured WHILE programming language is stored it is possible to represent it by a \textbf{tree} -- an intuitive example of this is the ability to fold code in environments such as Eclipse \cite{_folding_????}. The reason this is possible is that the body of a WHILE loop or an IF body can not jump to any other instruction outside its parent WHILE/IF instruction. Even though it is possible to model WHILEs and IFs using GOTO commands, the GOTO would also allow ``forbidden jumps'' (for example to the same command), therefore necessitating a graph data structure for the algorithm's representation.

Since graphs can not generally be represented by trees, to be able to transform a GOTO representation to a WHILE representation, constraints have been introduced in vIDE on the flowcharts that can be drawn in it:
\begin{itemize}
  \item  loops can only be drawn to go back to the last branch predecessor (of the same branching depth)
  \item  if a branch is not a loop its children blocks must join at one point (they need to share a common successor) on the same branching depth
\end{itemize}

Using these two constraints we effectively mask a WHILE language using a GOTO language, without the fear of not being able to convert it to a tree representation.

\subsection{The transformation algorithm}
 
Once we set the constraints, we can define an algorithm for transforming a constrained GOTO algorithm (illustrated in Fig. \ref{vIDE-class_diagram}) into a WHILE algorithm.

The GOTO-to-WHILE algorithm is essentially a depth first search (DFS) that recursively processes the nodes of a GOTO algorithm collecting single instructions to blocks and translating branches to WHILE or IF commands using processed node ``colouring'' (see \cite{cormen_introduction_2009} for DFS and similar graph traversal algorithms).

\begin{algorithm}
\caption{The GOTO-to-WHILE algorithm}
\label{alg:g2w}

\begin{algorithmic}
\WHILE{instruction exists}
	\IF{instruction is a block}
		\STATE nothing special
	\ENDIF
	\IF{instruction is a branch}
    		\STATE trueChild $\gets$ true child of the branch
    		\IF{G2W(trueChild) found processed command}
        		\STATE translate instruction to WHILE
        	\ENDIF
      	\IF{G2W(trueChild) reached program end}
      		\STATE translate instruction to IF
      	\ENDIF
	\ENDIF
	\STATE mark instruction processed
    \STATE instruction $\gets$ next instruction
\ENDWHILE
\end{algorithmic}
\end{algorithm}

The basic idea can be seen in Algorithm \ref{alg:g2w}. It is meant to be called as \emph{G2W(instruction)}, where the first instruction of the main block of the algorithm should be provided to the initial function call.
 
Note that \textit{G2W(trueChild)} represents processing child instructions of a branch in the case of the condition being met -- going further in depth recursively on the true side. Also, this is only a sketch of the ``interesting'' parts of the algorithm. In an actual implementation:
\begin{itemize}
  \item  All the constraints need to be checked and any breeches communicated to the user. 
  \item  Also, resolving the IF transformation is a bit more complex and requires tracing the condition being false branch as well and fixing the tree afterwards (because at first all the successor instructions would be added to IF as children -- an intersection, i.e. the end of the IF instruction, can only be found after tracing the second branch -- false).
  \item  The last omitted aspect is the communication -- a lot of information needs to be passed as arguments to recursive function calls or using a stack to identify the true state of the graph.
\end{itemize}

These details were skipped, since they would complicate the pseudo-code a lot and they can be implemented quite routinely.

\section{ Model-to-code transformation}
\label{sec:model-code}
 
After we have the algorithm represented in a tree-like structure of a structured WHILE language, the code generation is very simple and consists of:
\begin{enumerate}
  \item  a DFS through the tree
  \item  translating instruction objects to strings in the goal language syntax (Python in vIDE's example) on a one-to-one mapping basis.
\end{enumerate}

The result of this last procedure is a textual file containing a classical program. This program contains the steps equivalent to the flowchart used to generate it, specified through the goal language syntax. A computer can then execute this program and show the user the outcome of his flowchart.


\chapter{Visually defined procedural scene generation} 
\label{sec:procedural}


\ifpdf
    \graphicspath{{3/figures/PNG/}{3/figures/PDF/}{3/figures/}}
\else
    \graphicspath{{3/figures/EPS/}{3/figures/}}
\fi

The approach to procedural scene generation presented in this thesis is to have a library that exposes an API to the programmer for:
\begin{enumerate}
  \item easy object placement inside a scene
  \item information about the important markers in the scene for meaningful placement
  \item a simple set of randomness-providing functions for a more organic look and non-determinism in the generation
\end{enumerate}

A library was implemented to follow this approach and its name is \emph{libprocedural}. This implementation will be used as a model for studying the methods one would use for visual definition of procedures for automatic scene generation.

Like it was explained in Section \ref{sec:introduction.procedural}, most of the existing approaches rely on formal grammars for defining a scene. The novel approach presented here is to use a visual programming language in this process.

The library was designed in a way to suit the context of visual programming, which will be seen in the next Section \ref{sec:system_architecture.libprocedural}. This ``visual readiness'' guideline justifies some decisions in the library design (such as exposing parts of the library to the scripting language Python), which may otherwise seem redundant. Because of this, more emphasis will be put on the design of the library itself -- its deployment, API, exposed modules etc.

After that, in Section \ref{sec:libprocedural.api}, the library API is presented. This API is what the user sees through the visual programming language.

The last Section \ref{sec:libprocedural.modules} describes the actual algorithms used for procedural generation in each of the individual modules of the procedural generation library that is exposed to the user. An important factor while designing these modules was to let them \textbf{maximize the user's expressiveness in the scene description, while not straying too far off into low-level programming} of computer graphics.

The choice of modules to implement as part of this thesis was restricted to the domain of urban modeling -- procedural city generation, as a use-case example. It could analogously be expanded to satisfy a different domain or a broader problem of more general scene generation by following similar design patterns and guidelines.

\section{System architecture -- libprocedural}
\label{sec:system_architecture.libprocedural}

The idea for implementing procedural generation of scenes is to use an existing graphics engine, Ogre \cite{_ogre_????}, capable of importing models, placing them in a virtual scene etc., add additional higher-level logic on top of the graphics engine for procedural generation and encapsulate it in a library -- libprocedural. The user would then get presented with libprocedural's API for scene modeling, which would be on a higher level of abstraction than Ogre, and would be able to quickly write a definition of a scene -- by writing a program which interfaces with libprocedural, i.e. instantiates its objects and calls its methods. Two main qualities are required from the API:
\begin{enumerate}
  \item \textbf{it should be expressive} -- the user should be able to build complete scenes using them in a relatively small amount of code
  \item \textbf{the syntax should be clean and simple} -- since the end goal is to be able to visually define the procedure for scene generation (as described in Section \ref{sec:goals.procedural}) and this API will be the exposed in a VPL.
\end{enumerate}

Since vIDE, the goal VPL, generates code in Python from the drawn flowchart, as mentioned in Section \ref{sec:system_architecture}, it would be desireable that the procedural generation library could be accessed from Python as well. The library libprocedural is, on the other hand, programmed in \C++ \cite{stroustrup_c++_2000} because of:
\begin{itemize}
  \item \textbf{speed and efficiency}
  \item \textbf{interfacing with Ogre}, the underlying graphics library, also written in \C++.
\end{itemize}

To overcome these challenges, Python bindings were created for the \C++ library libprocedural. That way the library can be used from vIDE through Python. This goes in line with one of the already mentioned requirements put on the API -- a clean and simple syntax. An additional bonus is that the core logic remains in \C++, thus still staying fast \footnote{Python is on a higher level of abstraction than C/\C++ and is thus slower, which is one of the reasons it is not used as much as C, as can be seen in Fig. \ref{programming_language_popularity}. A rising curve can be seen, though, since Python interpreters and JIT compilers such as PyPy get more efficient and the language gets broader adoption.} and being able to delegate the graphics-related work to the Ogre engine.

In order for the library to be accessible from Python, it is built as a shared object (or a dynamically-linked library, in other terms).

For clarity's sake let us first give a simple description of what libprocedural physically is.

\subsection{A short description of libprocedural}

Because libprocedural's internal workings are a bit complicated, let's first try and give a short, down-to-earth description of what it is.

Looking from the filesystem side, libprocedural is a collection of \C++ classes -- headers (*.h) and implementation (*.cpp) files. Some of the files that are inside this library are:
\begin{itemize}
  \item \verb#SceneCreator.h#
  \item \verb#SceneCreator.cpp#
  \item \verb#BuildingGenerator.h#
  \item \verb#BuildingGenerator.cpp#
  \item \verb#GlobalData.h#
  \item \verb#GlobalData.cpp#
  \item  \ldots
\end{itemize}

These \C++ files are compiled to a shared library, libprocedural -- it can be linked by another \C++ program at runtime.

The purpose of libprocedural's classes is to manipulate an Ogre scene -- add new models into it. They can be instantiated and their methods called as per the class diagram shown in Fig. \ref{libprocedural-class_diagram}. Mostly they can be operated by creating an instance of a generator class (for example a building generator) and telling it to generate something (e.g. a building) on a location provided through the parameters. One generator class can be used many time. More on the topic of using libprocedural's API will be said later in Section \ref{sec:libprocedural.api.analysis}.

The entry class for the whole library is the SceneCreator class. This class has to be instantiated in an Ogre application and a scene manager object provided to it -- this object gets stored by libprocedural in a GlobalData singleton class for later use by other scene-manipulating classes. The SceneCreator has a \emph{runProcedure} method, which when invoked starts the procedural scene creation process. The scene creation occures according to a file named ``build\_scene.py" that is required inside the working directory (denominated by a ``.'' on Unix).

The SceneCreator class doesn't call other classes (BuildingGenerator etc.) directly. There is a Python layer in the middle of them, acting as an intermediary. The code executed in this Python layer is specified in the build\_scene.py file. This extra layer exists so that the procedure that describes how the scene is going to be built isn't written in \C++, but in Python -- a higher-level language, offering clearer and simpler syntax.

The library dependencies of libprocedural are as follows:
\begin{itemize}
  \item libogre -- the library containing the Ogre open source graphics engine
  \item libpython -- the Python interpreter is required so that the scene description, which is written in Python, can be run
  \item libboost\_python -- the Boost.Python library is built 
\end{itemize}

The file dependencies of libprocedural are:
\begin{itemize}
  \item \verb#build_scene.py# -- the steps of the procedure according to which the scene gets created
\end{itemize}

The build\_scene.py can be provided to libprocedural at runtime, it doesn't have to be compiled with the library. This allows for easy editing of the way the libprocedural creates the scene.

\subsection{Interfacing between libprocedural and Python}
\label{sec:libprocedural.python_interface}

One of the design decisions concerning libprocedural in a framework of visually defining procedures for automatic scene generation was to have \textbf{a Python interface for the library to make it more suitable to the visual programming context} from which it was to be used. Let us first examine how this interfacing is performed, since it determines the architecture of the remaining elements of the library and its usage a lot.

An illustration of this design pattern can be seen inside the module libprocedural in Fig. \ref{libprocedural-deployment}. In short -- a Python caller \C++ object invokes a Python interpreter that runs an arbitrary Python script and this script can in turn access the same libprocedural \C++ API again through the Python bindings. It is two-way communication. Let us now break it into smaller pieces and explain them one by one.

\subsubsection{Python $\rightarrow$ \C++}

To wrap the \C++ classes of our library and make them accessible from Python, Boost.Python, part of the highly renowned Boost \cite{karlsson_beyond_2005} \C++ library was used after comparing it to some of the other options available\footnote{namely SWIG}. Boost.Python can be used to wrap a set of \C++ classes as a Python module containing equivalent Python classes. One of the benefits of Boost.Python is that it offers good support for some of the trickier \C++ constructs such as inheritance, virtual classes/methods, default arguments (which will be important in libprocedural, as will be seen in Section \ref{sec:libprocedural.api.simple_by_default}) etc.

A set of classes is chosen to be exposed to Python. In case of libprocedural the classes representing APIs for invoking the abstract methods presented in Section \ref{sec:libprocedural.modules} and they are shown as a class diagram in Fig. \ref{libprocedural-class_diagram} (other helper classes exist, of course, but they are used in the implementation only, so the user is not burdened with having them exposed to him and they are not shown in the class diagram either). A description of how these classes are translated to Python is written using the Boost.Python library is required.

\subsubsection{Automatically creating the bindings}

It is possible to manually describe the Python bindings using the Boost.Python \C++ library, but this can be tiresome, the description has to be synced to the library itself (if something is changed in the \C++ API of the library, the Boost.Python description of the bindings has to be updated) and therefore an error-prone process.

To make this simpler, Py++ \cite{yakovenko_c++_????}, a tool tool written in Python, exists for automatically parsing a \C++ project and generating the Boost.Python bindings description. It can be invoked from a GUI or (for more fine-grained control) from a Python script. In the end it outputs a *.cpp file with a description of the Boost.Python bindings.

The library project can now be built with the bindings description included and libprocedural can be used from withing a Python interpreter as a module called procedural, as long as there is a procedural.so file (a file link to the original library itself, for example) in the working path.

\subsubsection{\C++ $\rightarrow$ Python}

It will be described in more details later that it is also required to be able to call the python interpreter from \C++ to execute an arbitrary Python script (the one that gets generated from the user's flowchart in vIDE).

This is achieved by including libpython -- the library containing the implementation of the Python interpreter, in a \C++ project (for example in libprocedural). After this, a function in an arbitrary Python script can be called by providing its location through the interpreter's API.

\subsection{Ogre -- the underlying graphics library}
\label{sec:librpocedural.ogre}

Ogre \cite{_ogre_????} is one of the most popular open source computer graphics libraries. It is a scene-oriented, flexible 3D engine written in \C++ designed for the development of applications utilising hardware-accelerated 3D graphics.

It can be used in a wide variety of domaincs:
\begin{itemize}
  \item games
  \item simulations
  \item business applications
  \item anything else requiring 3D graphics
\end{itemize}

More about Ogre can be read in \cite{kerger_ogre_2010}.

Due to its popularity and simplicity, Ogre was used as the underlaying graphics library in libprocedural. The idea is to delegate the scene creation task to libprocedural, that will then populate it with 3D models, using Ogre for their placement and manipulation. Basically, any application that uses Ogre can be extended with libprocedural in the part of the program where a scene has to be drawn. After that, Ogre can be used to further hone certain aspects of the scene and do the rest of the work -- add interaction, animation and similar capabilities.

\subsection{Deployment of libprocedural alongside vIDE}
\label{sec:libprocedural.deployment}

Let us now examine how could libprocedural be deployed in a larger framework -- with vIDE as a visual programming GUI and a target Ogre application that needs a procedurally generated scene drawn inside it. Fig. \ref{libprocedural-deployment} sketches the interactions between these components.

\figuremacroW{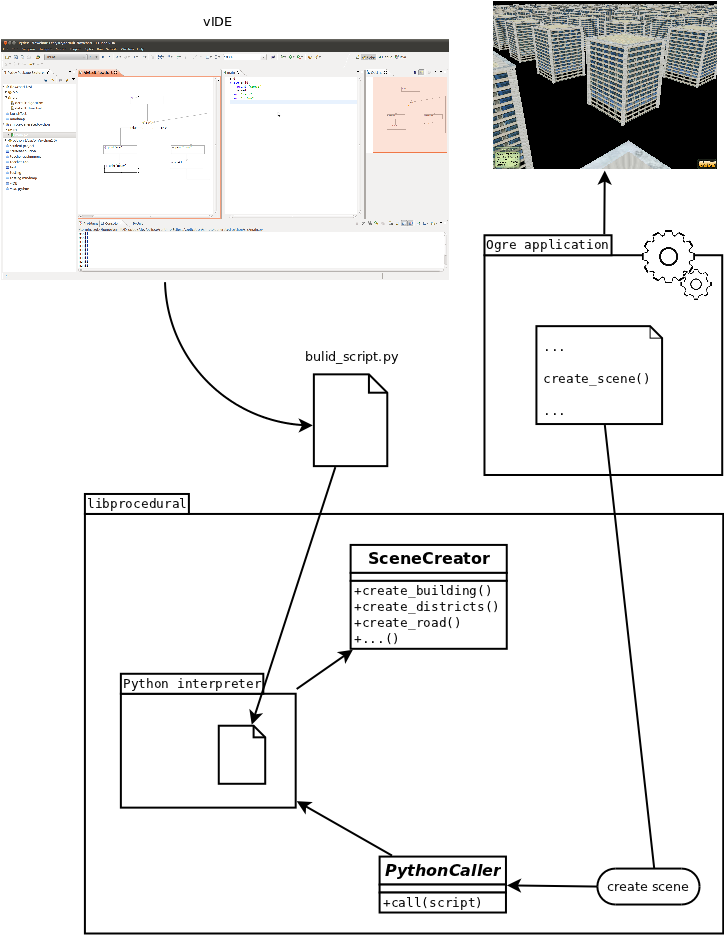}{Deployment of the libprocedural library in a wider context}{Components of libprocedural responsible for coupling with an Ogre application (used for rendering the scene) and vIDE (used for visually defining the procedure)}{0.9}

\subsubsection{Including libprocedural}

Let us suppose a user wants to deploy libprocedural in his application. The prerequisite is a target Ogre graphical application (a very simple example will do, such as the Tutorial Framework in \cite{kerger_ogre_2010}). Since libprocedural is a shared object, the inclusion to the Ogre application goes the same as for all the other such libraries -- it consists of linking to the libprocedural.so file and the header files. To use the library, the user would then have to:
\begin{enumerate}
  \item add the path to the libprocedural \C++ header files to his include path and the path to libprocedural.so file to his libraries path (and explicitly include the library itself) when he is compiling his Ogre application
  \item contain the path to the libprocedural.so file in the LD\_LIBRARY\_PATH environment variable for dynamic linking to work
\end{enumerate}

If the library headers were located in \emph{/usr/local/include} and the shared object file in \emph{/usr/local/lib} (standard Unix locations), a GCC compiler command used to build a program that includes vIDE might be:

\verb#g++ -I/usr/local/include/procedural main.cpp#

and the resulting executable can be run (since \emph{/usr/local/lib} is already one of the default linker search paths -- if it was not, it would be neccessary to add it first using standard BASH commands).

\subsubsection{Using libprocedural}

The library, libprocedural, is written in \C++ and at first needs to be accessed from an Ogre application as such, since Ogre is also written in \C++. Because the library is a shared object, the Ogre application can simply import it and initiate the creation of the scene via its \C++ API. Let us take a look at how would this be achieved\ldots

Once the library is included in the Ogre project, it can be used inside the part of the program where we want the scene to be created. This is achieved by first including the header:

\begin{lstlisting}{language=C++ centering}
#include "SceneCreator.h"
\end{lstlisting}

Once the Ogre application wishes to draw a scene, it can instruct libprocedural to do so and provide it with the Ogre scene manager object, so that it knows where to place the graphical objects it creates -- this action is represented as the connection between the Ogre application and libprocedural in Fig. \ref{libprocedural-deployment}. The entry class for libprocedural is SceneCreator, as can be seen in Fig. \ref{libprocedural-class_diagram}. The scene manager object is given to a SceneCreator through a constructor and gets stored in libprocedural's GlobalData singleton class to be accessible at any time and hide this low-level data away from the user. After we have a SceneCreator it is just a matter of calling the main procedure that does -- procedural scene generation:

\begin{lstlisting}{language=C++ centering}
SceneCreator creator(mSceneMgr);
creator.runProcedure();
\end{lstlisting}

Note that the scene can additionally be ``manually'' processed inside the Ogre application at this point -- for example shadows or special effects such as fog can be added. All the elements of the scene created by libprocedural can be accessed through the scene manager object and edited if necessary.

It remains to be seen how does the library knows what exactly should it generate\ldots

\subsubsection{Connecting to the scene description in Python}

The program has now entered the \C++ libprocedural library. This part of is presented inside the libprocedural module in Fig. \ref{libprocedural-deployment}. Now the library needs a description of the scene, so that it knows what and how to place in it. To achieve this, libprocedural loads the build\_scene.py file (which is expected to be inside the working directory) and starts a Python interpreter which executes the script.

From now on, the program runs in Python (well, the Python interpreter to be precise, but executes commands specified in the build\_scene.py file). The build\_scene.py is basically a high-level scene description made using libprocedural's Python API to call the individual procedural generation modules (such as the building, district and details generator classes, which will be explained later).

\subsubsection{Placing graphical elements in a scene}

In order to be able to invoke this API, the build\_scene.py script must load the libprocedural shared object. Note that we are linking it twice -- first when the Ogre application created the SceneCreator object and now once again when we want to access the individual procedural modules from Python (BuildingGenerator, DetailsGenerator\ldots) By calling functions available in libprocedural, graphical objects get placed in the scene manager of the Ogre application that started the whole process. The scene manager is the same one that was provided during library initialization\footnote{because the object has not been deleted yet -- the Ogre application is lower on the call stack and is waiting for libprocedural to finish its generation} and stored in a singleton class -- so that the user does not get bothered by having to know about it.

The scene description is now used to create the procedural scene. When all of the functions finish, the program control returns to the Ogre applications which started everything.

\subsubsection{Accessing the scene description from vIDE}

It was explained how to generate a scene from a definition stored in the build\_scene.py Python script. This script should be situated inside the working directory -- from where the Ogre application is executed. In order to create the procedural scene definition visually, a user also needs a vIDE project, aside from the \C++ Ogre application project which has libprocedural linked in it. This vIDE project can then be configured to generate a Python script from the flowchart that the user draws and store it inside a build\_scene.py file inside the Ogre application folder, When invoked, libprocedural can then access it. This is represented as the arrow build\_script.py file traveling from vIDE to libprocedural in Fig. \ref{libprocedural-deployment}.

Once this is done, the user interacts with vIDE, creating blocks of code and branches, connecting them the way he likes, and once he runs it generates the Python script directly inside the Ogre application project, rewriting its old contents. Following this, \textbf{the Ogre application can be started to examine the resulting scene that the user defined -- by editing the flowchart}.

The place where Python interface comes in is inside individual blocks and branches -- where the user writes individual commans. Instead of being forced to write in \C++ where he would need more commands, complex syntax, memory management etc., the user simply writes commands in Python that corespond to libprocedural's Python API.

\section{The library's API}
\label{sec:libprocedural.api}
The library's exposed classes and the programming entry point -- the SceneCreator class (that was already explained in Section \ref{sec:libprocedural.deployment}), can be seen in Fig. \ref{libprocedural-class_diagram}.

This API can be used directly in \C++ or from Python, by importing the \emph{procedural} Python module (see Section \ref{sec:libprocedural.deployment}).

\subsection{Simple by default, configurable when necessary}
\label{sec:libprocedural.api.simple_by_default}

\ldots was the main philosophy behind the design of individual classes that were to act as an interface to libprocedural.

This was mainly achieved by using \textbf{default argument values for methods so that the user is only obliged to enter the essential ones} at first, but is allowed to configure the others if he wishes to do so -- \textbf{keeping it simple, while retaining expressiveness}.

To show this on an example -- the \emph{generate} method of a building generator takes only the location of a building as the necessary input (the X and Y coordinates of the building center). The user can, if he wishes to, add additional arguments such as the building height, width\ldots

\subsection{The classes exposed to Python}
\label{sec:libprocedural.api.exposed}

All the classes shown in Fig. \ref{libprocedural-class_diagram} are exposed to Python, except for the SceneCreator (highlighted in the diagram, it was added since it is the entry point for an application that wants to use the library and it may therefore concern a potential user). Furthermore, they are exposed to vIDE through Python for the visual definition task -- when a user is drawing his flowchart in vIDE, he can use the exposed classes inside the blocks.

\figuremacroW{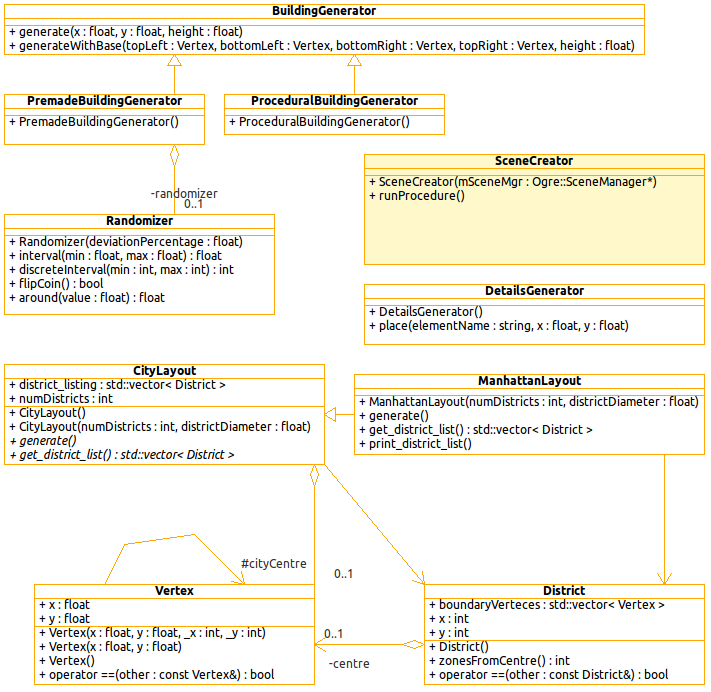}{A class diagram of the libprocedural API}{All the important classes in libprocedural that are used as an API for procedural generation of certain parts of the scene and the ``starter'' SceneCreator class}{0.9}

\subsection{Analysis of the API}
\label{sec:libprocedural.api.analysis}

Let us give a quick run-down of the API. Only the interface will be discussed -- more detailed description of what happens inside the classes can be found in Section \ref{sec:libprocedural.modules}

\subsubsection{Creating the city layout}

The CityLayout class can be used to create districts in a city. Currently, only a Manhattan-like layout can be generated. The optional arguments to the constructor are the number of districts to create and the district diameter. After initialization, the user has to call the \emph{generate} method first.

Following this the user can iterate over the district list using \emph{get\_district\_list()} and use the district objects' boundary vertices (iterating over them too, analogously) to place objects (for example buildings, trees\ldots) relative to them.

\subsubsection{Adding buildings}

Buildings can be added by instantiating a BuildingGenerator and then calling the \emph{generate} method. Like mentioned in Section \ref{sec:libprocedural.api.simple_by_default}, this method has the location as the obligatory arguments and the height, width etc. can optionally be specified.

Note that BuildingGenerator is a virtual class, instead one of its two subclasses -- a premade and a procedural building, must be created instead. The difference between how these two types of buildings are created will be discussed in Section \ref{sec:libprocedural.modules}.

\subsubsection{Adding details}

Again, a DetailsGenerator has to be instantiated. The main method this time is \emph{place} and it takes the name of the element to be placed and the location.

\subsubsection{Adding randomness}
\label{sec:libprocedural.api.classes.randomness}

A Randomness class exists to make all the procedurally generated content more natural. After the class is first instantiated, a random number generator is seeded. An optional argument is the scatter factor, which defaults to 0.1 or $10\%$, explained later. The same instance can be used from then on. A couple of methods exist here that can be used in all the other modules to make scenes a bit more undeterministic:
\begin{itemize}
  \item \emph{interval(a, b)} -- returns a random decimal number between a and b
  \item  \emph{discreteInterval(a, b)} -- same as above, but rounds the number
  \item \emph{around(x)} -- returns a random decimal number between $x*(1-scatter)$ and $x*(1+scatter)$, or in other words if scatter is 0.1 we get a value $x \pm 10\%$
  \item \emph{flipCoin()} -- returns true or false with a $50\%$ chance
\end{itemize}

By using one of these methods, values which would normally be constant and deterministict will be different everytime a program is executed. This is how a user can achieve randomness in his procedureal scene description.

\section{Methods used for procedural generation}
\label{sec:libprocedural.modules}

The methods presented here can be used to fulfill the proof-of-concept domain that was set on procedural generation as part of this research -- urban modeling.

Techniques on how to deal with the low-level graphical objects that are to be used as ``bricks'' are proposed, along with some concrete algorithms (for city layout generation and procedural building generation). Aditionally, a method of achieving randomness is presented that results in more convincing scenes and poses a creativity tool for defining unpredictable, non-repeating scenes that change on every encounter.

\subsection{Premade building blocks}

In order to create a procedural scene of some complexity, basic building blocks are needed. In an extreme case, they might be basic graphical primitives, but generating an urban scene using only them would not be an easy task.

The approach presented here is to use certain premade elements and then be able to combine them. These elements are meshes and textures created in Blender \cite{roosendaal_official_2004}, a 3D modeling program. They are on various levels of complexity, depending on what is required by certain method.

Generating premade buildings and details basically imports complete meshes made in Blender and allows the user to place them on desired locations and scale them. This is useful in creating a model of a city fast.

A bit more low-level approach is used with procedural buildings. Procedural buildings import only textures created in Blender of various elements that can be found on a building:
\begin{itemize}
  \item doors
  \item windows
  \item walls
  \item the roof
\end{itemize}

Subsequently, procedural buildings will compose basic graphical primitives such as rectangles to put together a form of a building and map textures to appropriate places.

\subsection{Building generation}
\label{sec:libprocedural.modules.building}

Buildings form the most noticeable aspect of the city so let us first go over how they can be generated.

\subsubsection{Premade building generation}

As mentioned previously, premade buildings simply import Blender models and place them on a certain location. It is possible to scale them a bit, but in extreme cases this distorts their appearance (windows and doors get streched and skewed along with the building itself).

A collection of premade buildings is available and each time a building is generated, one of the models is chosen at random and placed inside the scene so that it seems more convincing.

\subsubsection{Procedural building generation}

As already mentioned, procedural buildings load only textures from a preexisting source. The model itself is built dynamically (using the Ogre API).

\begin{algorithm}
\caption{The procedural building generation algorithm}
\label{alg:procedural.building}
\begin{algorithmic}
\STATE $number\ of\ floors \gets floor(height / 3.5)$
\STATE $i \gets 0$
\WHILE {$i < number\ of\ floors$}
	\STATE $i \gets i + 1$
	\STATE generate four rectangles on locations appropriate for walls
	\STATE map wall and window textures to the rectangles
	\IF {$i = 0$}
		\STATE map door texture to one rectangle
	\ENDIF
\ENDWHILE
\STATE generate the roof rectangle
\STATE map the roof texture to the rectangle
\end{algorithmic}
\end{algorithm}

The input parameters are the location of the building, its width and its height. The rough steps of the procedure for building generation are shown in Algorithm \ref{alg:procedural.building}.

The algorithm calculates the number of floors from the building height (and rounded using the appropriately-named mathematical \emph{floor} function) and builds each floor separately to fit the building base. A floor is generated by putting together walls (3D polygons) on adequate locations and mapping wall textures to them. Afterwards, window, door (only on the ground floor) and roof textures are added.

This results in buildings which scale much better to different dimensions, because the model is constructed in a way that perserves the dimensions of basic elements (doors, windows, floors) -- as it should be.

\subsection{Randomness methods}
\label{sec:libprocedural.modules.randomness}

The methods behind the Randomness module are pretty straightforward, using standard random number generation on a computer and will not be discussed specifically, beyond what is already said in Section \ref{sec:libprocedural.api.classes.randomness}.

It is important to note the consequence of having a randomness module in procedural generation -- this means that \textbf{the scene will be different every time it gets generated}. This might influence simple things such as the appearance of individual buildings, but it could also make a huge impact on the whole scene -- different procedures altogether could be used, depending on the outcome of a single \emph{flipCoin} method. This leaves a lot of room for creativity by enabling the development of unpredictable, dynamical applications.

\subsection{City layout generation}
\label{sec:libprocedural.modules.layout}

The first step when scaling the city to larger proportions would be to calculate some sort of a geometrical frame in which to place smaller entities. That way the whole area could be subdivided into smaller portions -- districts.

These districts are basically polygons that occupy a certain area of the 2D space. They are represented as a list of Vertices that form their boundary. A sketch of a layout can be seen in Fig. \ref{layout-sketch} -- it can be seen how various objects, such as buildings, roads, pavements etc. can be placed using boundary vertices as relative position markers.

\textbf{Districts also contain semantical information} -- for example they can be asked how far away they are from the city center. This kind of information would be useful to create different kind of procedurally generated content inside different districts, depending on their global position. For example higher buildings could be placed closer to the center.

\figuremacroW{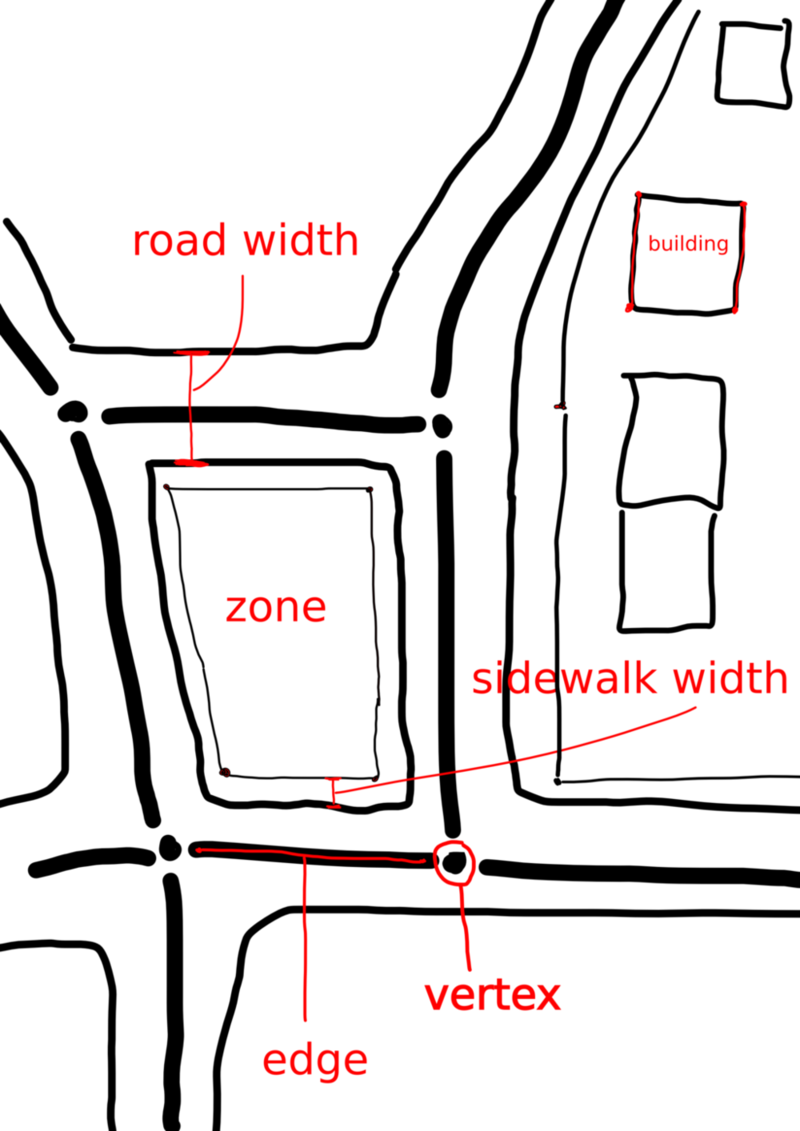}{Sketch of a district layout}{Boundary verteces of a district layout used for relative placement of other objects in the scene (buildings, roads, details \ldots)}{0.4}

\begin{algorithm}
\caption{The procedural Manhattan-like layout generation}
\label{alg:procedural.manhattan}
\begin{algorithmic}
\STATE $i \gets 0$
\STATE $xCoordinates \gets []$
\STATE $yCoordinates \gets []$
\STATE $dx \gets 0$
\STATE $dy \gets 0$
\STATE $lastX \gets 0$
\STATE $lastY \gets 0$
\WHILE{i $<$ numDistrictX}
	\STATE $i \gets i + 1$
	\STATE $lastX \gets lastX + dx$
	\STATE $lastY \gets lastY + dy$
	\STATE $xCoordinates.append(lastX)$
	\STATE $yCoordinates.append(lastY)$
	\STATE $dx \gets randomizer.around(districtDiameter)$
	\STATE $dy \gets randomizer.around(districtDiameter)$
\ENDWHILE
\STATE $i \gets 0$
\WHILE{i $<$ numDistrictX}
	\STATE $i \gets i + 1$
	\STATE $j \gets 0$
	\WHILE{j $<$ numDistrictY}
		\STATE $j \gets j + 1$
		\STATE $Vertex\ v_1 \gets Vertex(xCoordinates[i],\ yCoordinates[j])$
		\STATE $Vertex\ v_2 \gets Vertex(xCoordinates[i+1],\ yCoordinates[j])$
		\STATE $Vertex\ v_3 \gets Vertex(xCoordinates[i+1],\ yCoordinates[j+1])$
		\STATE $Vertex\ v_4 \gets Vertex(xCoordinates[i],\ yCoordinates[j+1])$
		\STATE create new district with vertices $(v_1,\ v_2,\ v_3,\ v_4)$
	\ENDWHILE
\ENDWHILE
\end{algorithmic}
\end{algorithm}

Districts are generated as a Manhattan-like grid of rectangules varying in size. The only minor obstacle is that they must fit together perfectly -- every two adjacent districts must have have the shared side of equal length. If we are provided with inputs: \emph{numDistrictsX} and \emph{numDistrictsY}, as the desired number of districts on the horizontal and vertical axis and \emph{districtDiameter}, quite obviously, as the desired district diameter (height or width), the layout can be generated as shown in Algorithm \ref{alg:procedural.manhattan}.

In the first loop of the algorithm we generate ``lines'' parallel to the X and Y axes at roughly uniform distances (using the \emph{around} method of the randomness module explained in Section \ref{sec:libprocedural.api.analysis} to get a bit more natural impression) and store their positions in two lists. In the second part, we go through all the (future) districts in a double loop. Inside the loop we instantiate four vertices on the appropriate positions (``two down, two up'') and create a district from them.

\subsection{Details placement}

Similarly to premade buildings, details are simply imported to the scene as premade models and placed. The collection of models can be expanded by creating new ones in Blender \cite{roosendaal_official_2004}.

Some of the available models are:
\begin{itemize}
  \item trees
  \item bus stations
  \item bicycle stands
  \item sign posts
  \item benches
  \item poles
\end{itemize}

Adding these objects to the scene creates convincing scenes, as they appear to be populated by everyday objects. Similarly to buildings, these objects could also be procedurally generated to ensure better scaling, memory preservation etc.


\chapter{Results} 
\label{sec:results}


\ifpdf
    \graphicspath{{4/figures/PNG/}{4/figures/PDF/}{4/figures/}}
\else
    \graphicspath{{4/figures/EPS/}{4/figures/}}
\fi


\section{Visual programming capabilities}
\label{sec:results.vide}

As a result of implementing all the presented methods, vIDE was built -- a VPL that can be installed as an Eclipse plug-in and used to draw flowcharts and generate Python code from them. In this section its capabilities will be examined.

\subsection{An example of vIDE usage -- code generation}

The example of the Euclid's algorithm drawn as a flowchart in vIDE is shown in Fig. \ref{Euclid-flowchart}.

\figuremacroW{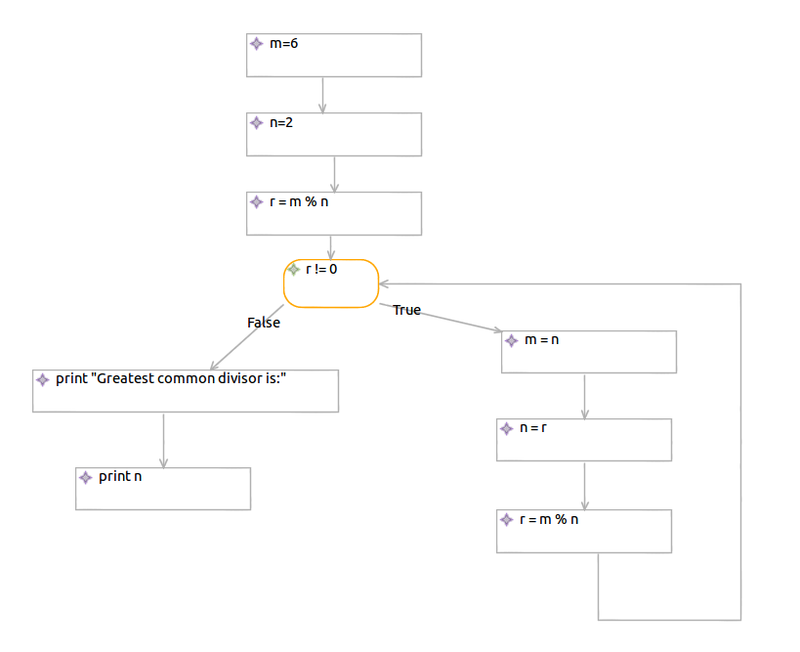}{Euclid's flowchart}{A flowchart of the Euclid's algorithm drawn in vIDE}{0.9}

From this flowchart the user can launch the code generation and a Python script is generated, equivalent to the algorithm defined in the flowchart:

\begin{center}
\begin{tabular}{c}
\begin{lstlisting}[language=Python]
m=6
n=2
r = m % n
while r != 0:
    m = n
    n = r
    r = m % n
print "Greatest common divisor is:"
print n
\end{lstlisting}
\end{tabular}
\end{center}

The branch in the flowchart was recognized by the GOTO-to-WHILE transformation algorithm to be a loop (because an instruction was pointing to its predecessor; note that a constraint was not breeched because it is the last branch) so a WHILE instruction in Python was generated.

\subsection{Constraint satisfaction in vIDE}

An example of automatic constraint satisfaction checking is implemented in vIDE, so that the user is not allowed to make a mistake in the first place. For example the user ca not connect a block to itself (a circular connection - it would result in an infinite loop), because the environment will not let him ``drop'' a connection on that position. This was implemented using the Object Constraint Language (OCL) \cite{_object_????}, a language for constraint definition which is integrated with GMF in a way that the checks are done in real time -- presenting the user with the currently relevant information.

The constraints defined in Section \ref{sec:constrained_goto}) will notify the user about a constraint breech only when he launches the flowchart-to-code transformation. This was not implemented in OCL, but in Java (because the check requires recursive function calls, which to our knowledge is not possible in OCL), as a part of the GOTO-to-WHILE transformation, so the notifications are not real-time.

\subsection{Knowledge inferring in vIDE}
\label{sec:knowledge_inferring}

One of the nice features of GMF is that it infers knowledge from the EMF model created to describe the diagram editor.

\figuremacroW{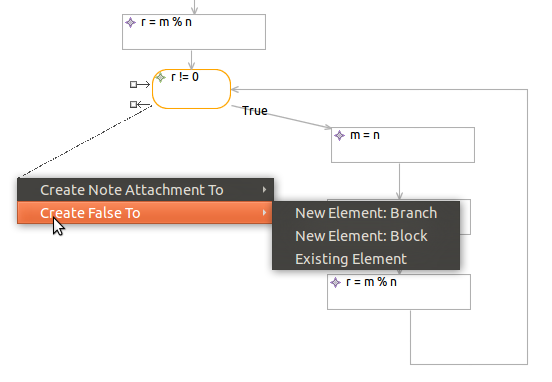}{Branch deduction}{Connection creation wth deduction (excess choices are trimmed)}{0.7}

An example in vIDE is that creating a connection from a branch for the first time asks the user whether it is the true or the false case, but when a true case is already present, the system deduces that there is no need in show this choice to the user as seen in Fig. \ref{branch-deduction}, resulting in only the semantically relevant information presented.


\section{Programmatically defining scenes with libprocedural}
\label{sec:results.libprocedural.programmatically}

The procedural methods presented in this thesis and implemented in the libprocedural library can be used from a classical programming language as well as from a visual programming language. This section will cover the programatical approach -- using libprocedural from \C++ natively and from Python through the library's Python interface as presented in Section \ref{sec:libprocedural.api}.

First, the native \C++ usage will be examined. Next, the same procedure specified in Python will be presented. By using these two different library interfaces in the goal of achieving the same behaviour, a good comparison between the user-friendliness of the two variations -- native \C++ and high-level Python, can be derived.

\subsection{Native usage of libprocedural}

To use libprocedural natively, one would have to merely import the library and instead of running the \emph{runProcedure} method simply use the various classes that are offered in it and that were described in Section \ref{sec:libprocedural.api.analysis}. All the code would go into a client Ogre application that uses the library when it needs to build a scene.

Let us say that we want to build a scene with a small number of physically separated districts and place buildings in a grid inside each of them. The \C++ code required to achieve this would state:

\begin{center}
\begin{tabular}{c}
\begin{lstlisting}[language=C++]
SceneCreator creator(mSceneMgr);

ManhattanLayout layout(9,2000);
layout.generate();

// scene parameters
float buildingDistance = 200;
float boundaryOffset = 300;
Randomizer randomizer(0.2);
PremadeBuildingGenerator building;

int i = 0;
while (i<layout.get_district_list().size()){
	// accessing data
	District district = layout.get_district_list().at(i);
	Vertex v1 = district.boundaryVerteces[0];
	Vertex v2 = district.boundaryVerteces[1];
	Vertex v4 = district.boundaryVerteces[3];
	
	float buildingX = v1.x + boundaryOffset;
	
	// moving on the X axis
	while (buildingX<v2.x-boundaryOffset){
		float buildingY = v1.y+boundaryOffset;
		// moving on the Y axis
		while (buildingY < v4.y-boundaryOffset){
			building.generate(buildingX, buildingY,
				randomizer.interval(50,100),
				randomizer.interval(40,50));
			buildingY +=
				randomizer.around(
				buildingDistance);
		}
		buildingX +=
			randomizer.around(
			buildingDistance);
	}
   ++i;
}
\end{lstlisting}
\end{tabular}
\end{center}

The SceneCreator was created with an Ogre scene manager and after that the classes were normally instantiated using \C++ syntax. First a Manhattan layout was generated, then a couple of global city parameters were set (the distance between two buildings, distance from the district boundary, a randomizer scatter factor etc.)

\figuremacroW{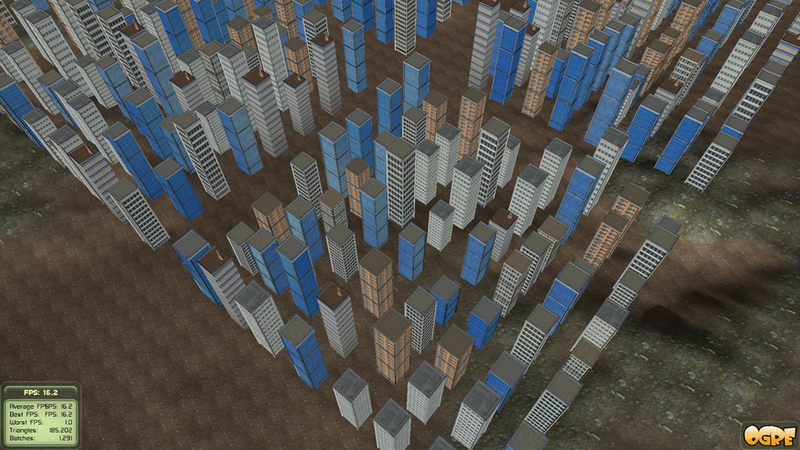}{Generated scene defined programmatically}{This scene was procedurally created from a program that uses libprocedural.}{0.9}

The next step is to iterate over all the districts and inside each district we move from one corner of the district's rectangular area to the other in two dimensions -- on the X and on the Y axis (using a double while-loop), with a district offset value so that there is some space left between adjacent districts. Buildings are generated inside the innermost while-loop at a predefined density (buildingDistance). The loop conditions check whether the boundary of the district was reached.

The outcome of running the program can be seen in Fig. \ref{programmatically}. The scene is viewed from above so that separate districts could be discerned. A distance between clusters of building is defined by a district offset parameter and the space between any two adjacent buildings is defined by the building distance parameter.

\subsection{Using libprocedural from Python}

To use libprocedural from Python, the library has to be imported like before and the \emph{runProcedure} method called. Additionally, like explained in Section \ref{sec:libprocedural.deployment}, a build\_scene.py Python script has to be provided. This time the procedure for building the scene is stated in this Python script and it uses the API described in Section \ref{sec:libprocedural.api.analysis}, but from Python. Only the code to start the procedure would go into a client Ogre application that uses the library when it needs to build a scene (see Section \ref{sec:libprocedural.deployment}). The rest of it gets delegated to the Python script and that's why only this script is of interest in this scenario.

If we were to build the same scene as in the last section -- a small number of physically separated districts with buildings in a grid inside each of them. The Python code in build\_scene.py required to achieve this would state:

\begin{center}
\begin{tabular}{c}
\begin{lstlisting}[language=Python]
from procedural import *

layout = ManhattanLayout(9,2000)
layout.generate()

# scene parameters
buildingDistance = 200
boundaryOffset = 300
randomizer = Randomizer(0.2)
building = PremadeBuildingGenerator()

i =0
while (i<len(layout.get_district_list())):
    # accessing data
    district = layout.get_district_list()[i]
    v1 = district.boundaryVerteces[0]
    v2 = district.boundaryVerteces[1]
    v4 = district.boundaryVerteces[3]
    
    buildingX = v1.x + boundaryOffset;
    # moving on the X axis
    while (buildingX<v2.x-boundaryOffset):
        buildingY = v1.y+boundaryOffset
        # moving on the Y axis
        while (buildingY < v4.y-boundaryOffset):
            building.generate(buildingX, buildingY,\
                                  randomizer.interval(50,100),\
                                  randomizer.interval(40,50))
            buildingY += randomizer.around(buildingDistance)
        buildingX += randomizer.around(buildingDistance)
    i = i + 1
\end{lstlisting}
\end{tabular}
\end{center}

The Python script got the program control when a SceneCreator object with an Ogre scene manager already exists and is accessible, so we can start dealing with the semantics of placing scene elements right away. A Manhattan layout is generated, then a couple of global city parameters are set -- note that their type doesn't need to be explicitly declared, it is inferred from the constant used to initialize them. We iterate over all the districts and place buildings inside its area relative to its boundary vertices.

After running this procedure, the exact same scene (if we disregard the non-determinism factor coming from the use of the randomness module) gets created as the one shown in Fig. \ref{programmatically}. The \C++ and the Python programs described above achieve the same results.

\subsection{Comparison of different interfaces}

Since the two programs presented above for programmatically defining a scene yield the same output shown in Fig. \ref{programmatically}, they are good subject for the comparison of the two languages and how they could be used to present libprocedural's capabilities to the user.

The logic of both the Python and the \C++ program is very similar, only with Python offering a visibly simpler syntax. The whole Python program is a bit shorter, due to these facts:
\begin{itemize}
  \item it doesn't require variable type declaration
  \item it doesn't require special block delimiters, such as \verb#{}#, which are used in \C++
  \item the district list and the boundary vertices list elements are all accessed using the \verb#[]# delimiter
  \item Python's \emph{len()} function can be used to count the numbers of elements in data structures
\end{itemize}

Most of these facts mostly come from a different programming language syntax and phylosophy employed in Python. Apart from the block delimiters, all of these aspects of a programming language would be visible to the user of the coresponding interface in a visual programming language as well.

\section{Visual procedural modeling capabilities}
\label{sec:results.procedural}

In this section it will be shown how one could create a scene by using vIDE's methods of visual programming (presented in Section \ref{sec:vIDE}) in combination with the methods of procedural modeling offered through libprocedural's Python interface (as detailed in Section \ref{sec:procedural}). First, a very simple example flowchart shall be given, onto which more complexity will gradually be added. Each time, the usage of an additional module will be presented.

\figuremacroW{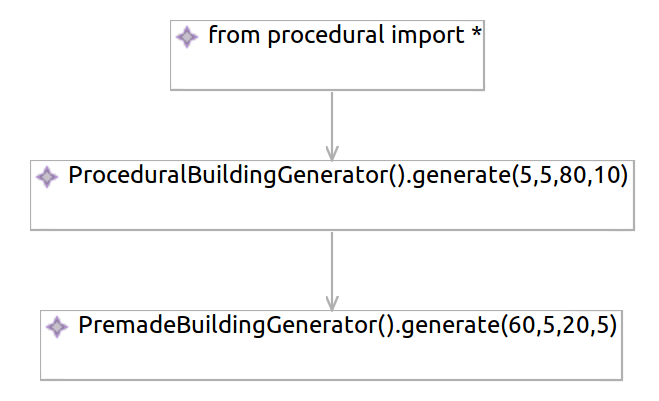}{Building generation flowchart}{This flowchart will generate two buildings -- one procedural and one premade.}{0.5}

The scene creation procedure is defined in vIDE and the resulting Python script generated and used by a simple Ogre application that uses libprocedural and the provided script to build the scene. The sky and the terrain are added manually in the Ogre application for completeness, but all other graphical objects in the scene come from libprocedural, following the procedure specified in a vIDE flowchart.

\subsection{Creating a building}

To compare the outcomes of creating a procedural and a premade building a flowchart for vIDE was drawn, as shown in Fig. \ref{procedural-flowchart}. It is a fairly simple example, where the procedural module is imported and two different building generators are instantiated and used.

\figuremacroW{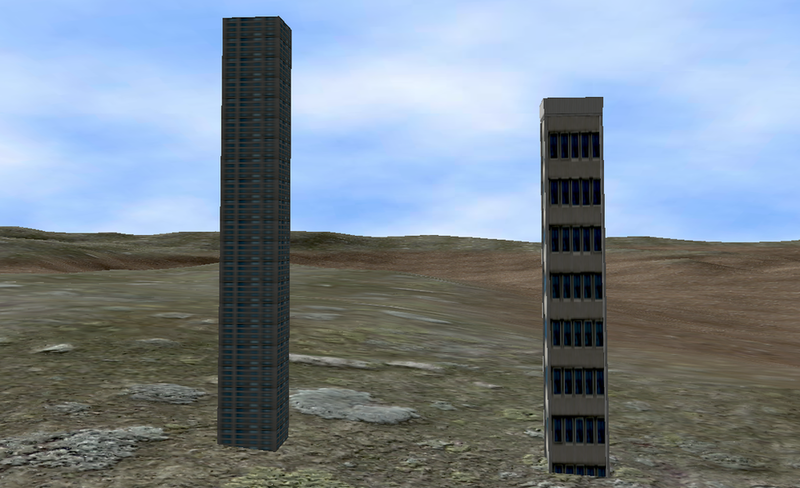}{Generated scene with a procedural and a premade building}{The procedural building (left) was composed from basic elements dynamically. The premade building (right) is a preloaded mesh scaled to fit the required size}{0.6}

The resulting scene can be seen in Fig. \ref{procedural}. The procedural building is composed dynamically by putting together walls and smaller textures as building blocks, while the premade building is a scaled mesh of a model previously created in Blender. It can be seen that the windows on the premade building are distorted, due to the stretching of the model, while the procedural building simply adds new floors when a higher building is required.

\figuremacroW{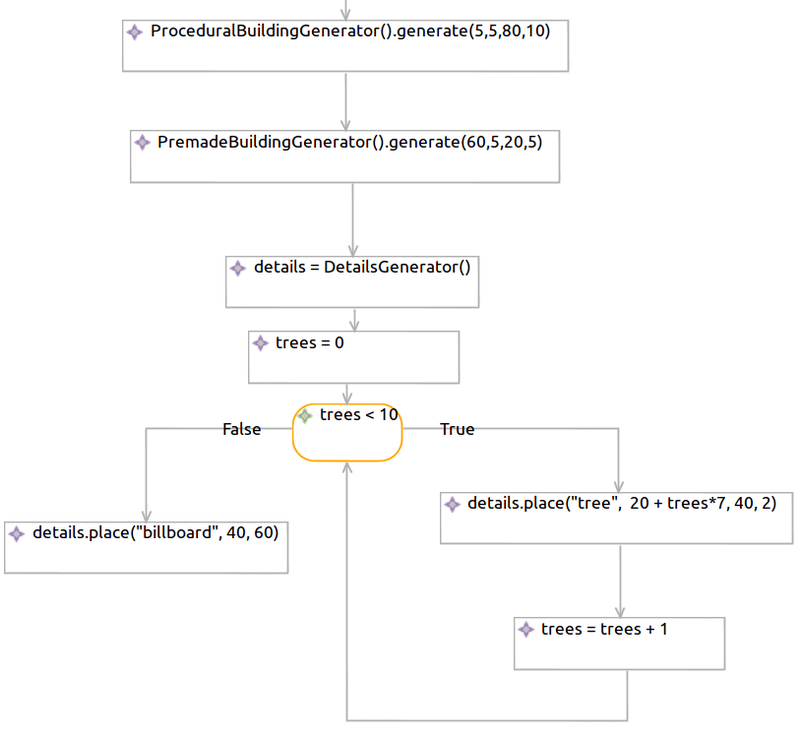}{Details generation flowchart}{This flowchart will add a row of trees to the scene  -- using a loop, and a billboard.}{0.9}

\subsection{Adding details}

The next step is to add some additional details to the scene. The flowchart from Fig. \ref{procedural-flowchart} was edited to add these details. The changes to the flowchart are shown in Fig. \ref{details-flowchart}. The new procedure creates a DetailsGenerator object and uses it to place a row of trees in the scene (using the VPL's loop pattern) and a billboard.

\figuremacroW{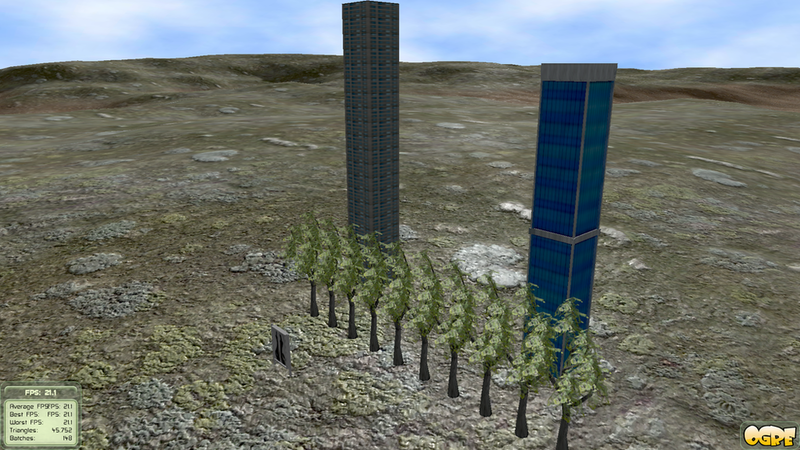}{Generated scene with details}{A row of trees and a billboard were added.}{0.7}

The resulting scene can be seen in Fig. \ref{details}. It can be seen that exactly 10 trees are created, as is defined in the flowchart through the loop condition. The coordinates of the trees are altered each time in the body of the loop using the expression $20+trees*7$, which results in a shift between every two trees in the scene.

\subsection{Generating a city layout}

An elegant way to enlarge the scene is by generating a Manhattan-like grid layout and then getting to the lower level of decorating a single district. We edit the previously explained flowchart from Fig. \ref{details-flowchart} to get a flowchart shown in Fig. \ref{districts-flowchart}. Basically, a 100-district layout is generate and we iterate through all of these districts using a loop. Inside the loop body the district's center is calculated and a single premade building placed here. Additionally, two trees are added to the edge of the district.

\figuremacroW{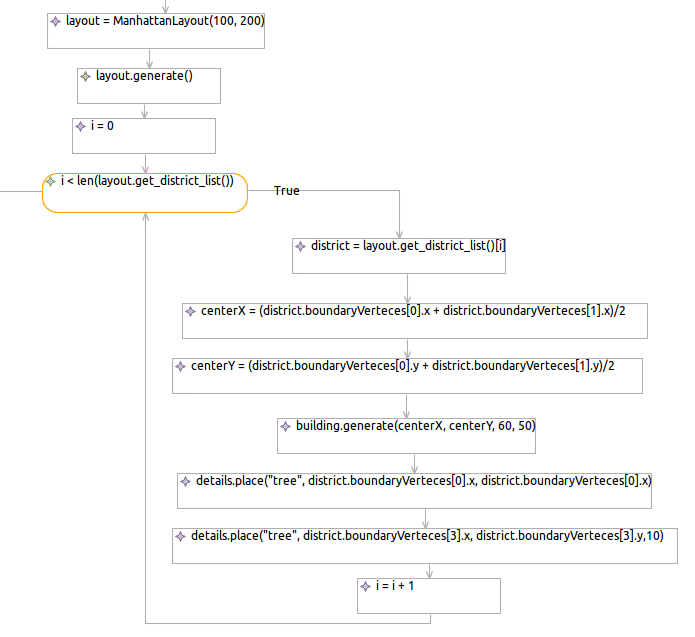}{Districts generation flowchart}{This flowchart specifies the creation of a Manhattan-like layout and the placement of buildings and trees inside of each of the districts.}{0.9}

Fig. \ref{districts} shows the outcome of this procedure. A two-dimensional city is created with trees in the midst of it.

\figuremacroW{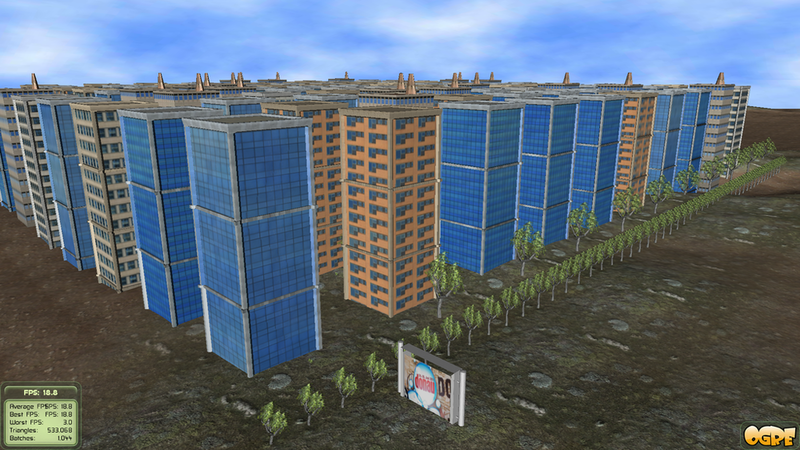}{Generated scene utilizing a layout}{Buildings and trees are placed relatively to an imaginary grid.}{0.7}

\subsection{Achieving non-determinism}

What is making the last scene, shown in Fig. \ref{districts} unconvincing is the way all the buildings and trees seem uniform, which seldom occurs in the real world. To fix this, randomness-inducing methods are added in the flowchart shown in Fig. \ref{randomness-flowchart}. Not much was added here -- simply an instance of the randomness class is used at certain places where exact values are used to describe objects (such as building heights, positions, tree heights\ldots) and its \emph{interval} and \emph{around} methods (described in Section \ref{sec:libprocedural.modules.randomness}) are used to scatter them a bit.

\figuremacroW{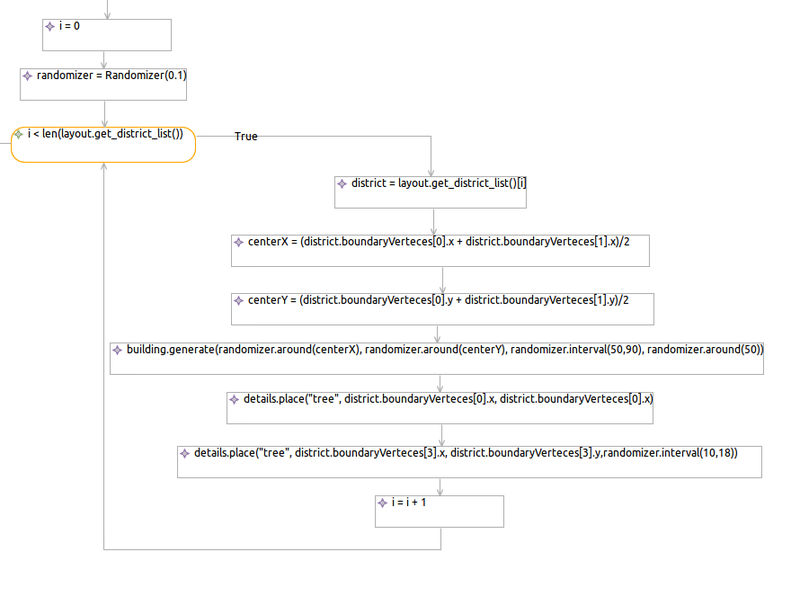}{Randomness-invoking flowchart}{This flowchart causes some unpredictability by setting building heights and locations and tree heights to random variables within a specified interval.}{0.9}

The outcome can be seen in Fig. \ref{randomness} and the improvement is evident over the deterministict scene shown in Fig. \ref{districts}, even though the change in the flowchart was not that big.

\figuremacroW{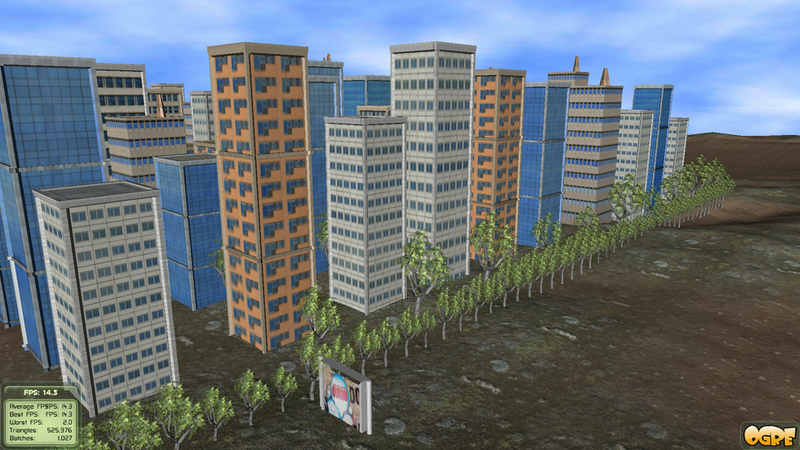}{A generated city with randomly-sized buldings}{This screenshot shows a city where buildings and trees look a bit more natural, as they are not all of the same height and the buildings are not positioned in a grid so strictly.}{0.9}





\chapter{Conclusion} 
\label{sec:conclusion}


\ifpdf
    \graphicspath{{5/figures/PNG/}{5/figures/PDF/}{5/figures/}}
\else
    \graphicspath{{5/figures/EPS/}{5/figures/}}
\fi


\section{Visual programming}

The proposed methods for building a VPL capable of flowchart editing and code generation have been shown to work on the example of vIDE. The VPL can be used for drawing and executing flowcharts comparable to \cite{_scratch_????}, while also giving the user the novel ability of generating clear, readable Python code, achieving the goals set in Section \ref{sec:goals.visual}. This is important, because it allows the users to grasp programming language syntax more easily and move to more complex, classical programming later (if needed). Such an application could find a lot of use in lowering the programming entry barrier and would be important for:

\begin{itemize}
  \item  didactic purposes -- teaching programming in an interactive and visual way
  \item  ease of usage -- making programming more accessible to people of other professions
\end{itemize}

Features such as knowledge inferring, described in Section \ref{sec:knowledge_inferring}, provide a great way of leveraging advanced GUI capabilities to provide a better programming environment. This goes in line with the general guidelines for creating VPLs stated in \cite{burnett_scaling_1995} and could be considered as a visual programming equivalent of context aware systems such as Mylyn \cite{_mylyn_????}.

Constraint definition using OCL enables real-time notifications. This seems to be a good way to utilize the advantages of VPLs -- namely a chance for preeminent notifications, as described in Section \ref{sec:strengths}. One of the problems is checking more complex constraints. Perhaps iterative transformation to the WHILE language model while the flowchart is being drawn would allow for more lively notifications.

\section{Procedural modeling through visual programming}

The presented visual programming language was successfully used in the domain of procedural generation using a set of methods exposed to the user through an API, fulfilling the goals stated in Section \ref{sec:goals.procedural}.

A proof-of-concept solution for the scenario of modeling urban scenes was presented and it offers:
\begin{itemize}
  \item building generation -- procedurally assembled and premade in Blender\cite{roosendaal_official_2004}
  \item details placement -- various objects can be added, such as trees, billboards, traffic signs, bus stops\ldots
  \item city layout generation -- a Manhattan-like grid can be calculated, enabling arrangement of objects relative to it 
  \item randomness -- all the parameters can be specified through intervals, rough values and random parts of the procedure could be executed even, resulting in more non-deterministic and interesting scenes
\end{itemize}

The flowcharts and resulting scenes presented in Section \ref{sec:results.procedural} show that the methods of visual programming and procedural generation can be used together to quite satisfying results -- especially considering the relatively small number of procedural generation methods utilized here in comparison to the very complex systems described in \cite{parish_procedural_2001,kelly_interactive_2007,wonka_instant_2003,mueller_procedural_2006}.

The procedural generation methods were designed in a way to suit the context of visual programming by masking the implementation complexity with a high-level interface of a scripting language Python \cite{rossum_introduction_2011} that gets presented inisde the VPL. In Section \ref{sec:results.libprocedural.programmatically} a side-by-side comparison shows that the usage of the library's native interface would burden the user with having to write more text inside a visual program to achieve the same functionality and we can therefore conclude that creating a high-level interface is a good choice.

The user's expressiveness in describing the scene was maximized, while not sacrificing simplicity by using interface design patterns such as \emph{default arguments}. This keeps true to the initial goals by using the ``Simple by default, configurable when necessary'' philosophy as described in Section \ref{sec:libprocedural.api.simple_by_default}. The user initially only has to provide the basic parameters, but is able to fine-hone his design later, which contributes to application scalability in the sense described in \cite{burnett_scaling_1995}.

The approach presented in this thesis might be found useful for:
\begin{itemize}
  \item rapid creation of content for entertainment purposes
  \item benchmarking graphical engines such as Ogre \cite{kerger_ogre_2010}
  \item education motivated by easy creation of interactive content, somewhere in the line of Scratch \cite{_scratch_????} and Alice \cite{_alice:_1995}.
\end{itemize}

A potential for experimental interactive applications is touched upon through the non-deterministic scene definitions discussed in Section \ref{sec:libprocedural.modules.randomness}. The scene can be different every time it gets generated. These dynamically built scenes could offer e.g. different experience when playing a computer game for the second time, or even when entering the same virtual room for the second time.

These and similar applications could greatly benefit from being able to visually describe procedures for automatic scene generation, using the methods presented here.







\begin{multicols}{2} 
\begin{tiny} 

\bibliographystyle{Latex/Classes/PhDbiblio-url2} 
\renewcommand{\bibname}{References} 

\bibliography{9_backmatter/references} 

\end{tiny}
\end{multicols}








\begin{abstractslong}    
With more and more digital media, especially in the field of virtual reality where detailed and convincing scenes are much required, procedural scene generation is a big helping tool for artists. A problem is that defining scene descriptions through these procedures usually requires a knowledge in formal language grammars, programming theory and manually editing textual files using a strict syntax, making it less intuitive to use. Luckily, graphical user interfaces has made a lot of tasks on computers easier to perform and out of the belief that creating computer programs can also be one of them, visual programming languages (VPLs) have emerged. The goal in VPLs is to shift more work from the programmer to the integrated development environment (IDE), making programming an user-friendlier task.

In this thesis, an approach of using a VPL for defining procedures that automatically generate virtual scenes is presented. The methods required to build a VPL are presented, including a novel method of generating readable code in a structured programming language. Also, the methods for achieving basic principles of VPLs will be shown -- suitable visual presentation of information and guiding the programmer in the right direction using constraints. On the other hand, procedural generation methods are presented in the context of visual programming -- adapting the application programming interface (API) of these methods to better serve the user. The main focus will be on the methods for urban modeling, such as building, city layout and details generation with random number generation used to create non-deterministic scenes.
  
These methods are demonstrated on an example of vIDE, a VPL based on the Eclipse IDE. The design of vIDE with respect to the Eclipse Graphical Modeling Framework (GMF) is described. The concept of a flowchart graphical notation is examined, its mapping to an algorithm data structure and the final conversion to a textual program (for example in the scripting language Python). The procedural generation functionality is encapsulated in a \C++ library libprocedural, which uses Ogre as a graphical engine. To make the interface between vIDE and libprocedural intuitive, high-level Python bindings were created.

\textit{Index terms} --- procedural scene generation, Visual programming, VPL, flowchart, programming language, GOTO, WHILE, vIDE, Eclipse, GMF, OCL, libprocedural, Ogre, \C++, Python.
\end{abstractslong}


\begin{abstractslongcro}


Uz sve više i više digitalnih sadržaja, posebno u području virtualne stvarnosti, gdje se traže uvjerljive scene, proceduralno generiranje scena je velika pomoć umjetnicima. Problem je jedino u tome što definiranje opisa scena ovakvim procedurama obično zahtjeva znanje formalnih jezičnih gramatika, teoriju programiranja i ručno uređivanje tekstualnih datoteka koristeći strogu sintaksu, što ih čini manje intuitivnima za korištenje. Na sreću, grafička korisnička sučelja su olakšala obavljanje mnogih radnji na računalu i radi uvjerenja da i izrada računalnih programa može biti jedna od njih, pojavili su se vizualni programski jezici (VPJ). Cilj u VPJ-icima je prebaciti više posla s programera na integriranu programsku okolinu (IDE), čineći programiranje korisniku-pristupačan zadatak.

U ovom je radu predstavljen pristup korištenja VPJ-a za definiranje procedura koje automatski generiraju virtualne scene.
Predstavljene su metode potrebne da se izradi VPJ, uključujući novu metodu generiranja čitkog koda u strukturnom programskom jeziku. Također, biti će pokazane metode za ostvarivanje osnovnih principa VPJ-ova -- prikladno vizualno predstavljanje informacija i vođenje programera u pravom smjeru koristeći ograničenja. S druge strane, predstavljene su metode za proceduralno generiranje u kontekstu vizualnog programiranja -- prilagodba programskog sučelja (API) ovih metoda kako bi bolje služile korisniku. Fokus će biti na metodama za urbano modeliranje, tj. generiranje zgrada, plana grada i detalja, uz korištenje generiranja nasumičnih brojeva kako bi se dobile nedeterminističke scene.
  
Ove su metode prikazane na primjeru vIDE-a, VPJ-a zasnovanog na Eclipse IDE-u. Opisan je dizajn vIDE-a s obzirom na Eclipse okolinu za grafičko modeliranje (GMF). Dan je pregled grafičke notacije dijagrama toka, njegovo mapiranje u strukturu podataka algoritma i konačno pretvaranje u tekstualni program (npr. u skriptnom jeziku Pythonu). Funkcionalnost proceduralno generiranja je zatvorena u \C++ biblioteku libprocedural, koja koristi Ogre kao grafički sustav. Kako bi se sučelje između vIDE-a i libprocedurala učinilo što intuitivinijim, veza prema jeziku visoke razine, Pythonu je napravljena.

\textit{Index terms} --- proceduralno generiranje scena, vizualno programiranje, VPJ, dijagram toka, programksi jezik, GOTO, WHILE, vIDE, Eclipse, GMF, OCL, libprocedural, Ogre, \C++, Python.
\end{abstractslongcro}


\end{document}